\documentclass[journal]{IEEEtran}
\usepackage{amsmath, amssymb}
\usepackage{tipa}
\usepackage{comment}
\usepackage{amsmath,amssymb}
\usepackage{cite}

\usepackage{textgreek}
\usepackage[linesnumbered,ruled,vlined]{algorithm2e}
\usepackage[caption=false,font=normalsize,labelfont=sf,textfont=sf]{subfig}
\usepackage{mathtools}  
\usepackage[table,xcdraw]{xcolor}
\begin{document}
\pagenumbering{gobble}
\title{\vspace{-7mm}Sparse Target Parameter and Channel Estimation in
MIMO-OOFDM-Aided Visible Light Integrated Sensing and Communication Systems}
\author{Shubham~Saxena,~\IEEEmembership{Graduate Student Member,~IEEE,} Garudadri~Mythreya, Suraj~Srivastava,~\IEEEmembership{Member,~IEEE,} Washim~Uddin~Mondal, and Aditya~K.~Jagannatham,~\IEEEmembership{Senior Member,~IEEE} \vspace{-8mm} 
\thanks{Shubham Saxena, Garudadri~Mythreya, Washim~Uddin~Mondal, and Aditya K. Jagannatham are with 
the Department of Electrical Engineering, Indian Institute of Technology 
Kanpur, Kanpur-$208016$, India (e-mail: shubs20@iitk.ac.in; mythreyag24@iitk.ac.in; wmondal@iitk.ac.in; adityaj@iitk.ac.in). Suraj Srivastava is with the 
Department of Electrical Engineering, Indian Institute of Technology 
Jodhpur, Jodhpur, Rajasthan $342030$, India (email: 
surajsri@iitj.ac.in).}}
%\date{\today}
\maketitle
\begin{abstract}
A unified framework is developed for parameter estimation (PE) and channel estimation (CE) in monostatic multiple-input multiple-output (MIMO) direct-current-biased optical orthogonal frequency division multiplexing (DCO-OFDM)-aided visible light integrated sensing and communication (VL-ISAC) systems. A common optical waveform supports sensing and downlink communication while accounting for target reflections, clutter, Doppler effects, and the communication channel. For sensing PE, one-dimensional ($1$D) and two-dimensional ($2$D) on-grid and off-grid sparse representations are formulated. The proposed $1$D off-grid Bayesian learning ($1$D-OffGBL) algorithm employs a joint delay-Doppler (DD) dictionary to exploit coupled DD information and achieve high estimation accuracy. In contrast, the proposed $2$D off-grid Bayesian learning ($2$D-OffGBL) algorithm sequentially estimates the Doppler and delay parameters, thereby avoiding a high-dimensional joint dictionary and reducing computational complexity. Both algorithms iteratively refine the continuous-valued delay and Doppler parameters to mitigate off-grid basis mismatch. Furthermore, $1$D and $2$D sparse channel representations and corresponding CE procedures are developed, followed by range and velocity estimation for both sensing models. Closed-form Bayesian Cramér-Rao bounds (BCRBs) are derived for PE, CE, range, and velocity estimation. Simulations demonstrate that the proposed algorithms outperform conventional sparse recovery schemes in terms of normalized mean-squared error (NMSE). In particular, $1$D-OffGBL achieves the highest estimation accuracy, whereas $2$D-OffGBL provides a favorable performance-complexity trade-off for practical MIMO DCO-OFDM VL-ISAC systems.
\end{abstract}

\vspace{-2mm}
\section{Introduction}
% \IEEEPARstart{I}{ntegrated} sensing and communication (ISAC) has emerged as a promising paradigm for intelligent environments requiring high-rate connectivity and situational awareness. By sharing time-frequency resources and a common transceiver architecture for communication and sensing (C\&S), ISAC can enhance spectral efficiency and reduce hardware redundancy \cite{9898900,10960484}. Although most ISAC research has focused on radio frequency (RF) systems, increasing traffic and RF-spectrum scarcity have encouraged exploration of optical bands. In this context, visible light (VL)-based ISAC (VL-ISAC) exploits the broad, unlicensed visible spectrum while offering low electromagnetic interference, enhanced security, energy efficiency, and compatibility with existing lighting infrastructure \cite{zhang2021evolution,6011734}.
\IEEEPARstart{I}{ntegrated} sensing and communication (ISAC) has emerged as a promising paradigm for modern intelligent environments requiring simultaneous high-rate connectivity and accurate situational awareness. By sharing time, frequency, and hardware resources between communication and sensing (C\&S), ISAC can improve spectral efficiency, reduce transceiver redundancy, and support a wide range of environment-aware applications \cite{9898900,10960484}. Although most existing ISAC research has focused on radio frequency (RF) systems, the rapid growth of mobile data traffic and persistent RF-spectrum scarcity have motivated its extension to the optical domain. In this context, visible light communication (VLC), operating over the largely unlicensed $400-800$ terahertz optical spectrum, offers a promising complementary technology. Visible light (VL)-based ISAC (VL-ISAC) combines optical C\&S using a common waveform and infrastructure. In addition to abundant bandwidth, VL-ISAC offers low electromagnetic interference, enhanced security, high energy efficiency, favorable signal-to-noise ratio (SNR) performance, and compatibility with existing lighting systems, supporting energy-efficient communication \cite{zhang2021evolution,6011734}.

VL-ISAC employs light-emitting diodes (LEDs) and photodiodes (PDs) to integrate illumination, communication, and sensing within a common infrastructure. Information is conveyed through intensity variations imperceptible to the human eye, while reflected optical signals provide target information, including range, velocity, and radar cross section (RCS). Since both functions use the same optical carrier, their integration improves resource utilization and environmental awareness. Multiple LEDs and PDs can also form a multiple-input multiple-output (MIMO) architecture, providing spatial diversity, sensing coverage, and observations for reliable target parameter estimation (PE) and channel estimation (CE).

VL channels generally comprise a dominant line-of-sight (LoS) component and multiple non-LoS (NLoS) paths generated by environmental reflections \cite{schulze2016frequency,zhou2014impact}. Higher-order reflections may contribute appreciably and should therefore be incorporated into accurate channel models \cite{zhou2014impact,saxena2023sparse}. The resulting multipath propagation introduces temporal dispersion and inter-symbol interference (ISI). Optical orthogonal frequency division multiplexing (O-OFDM) is well suited to mitigating these effects because of its spectral efficiency and robustness against multipath-induced ISI \cite{zhang2021evolution,saxena2023sparse}. Since intensity modulation and direct detection (IM/DD) requires a real-valued and non-negative waveform, direct-current-biased O-OFDM (DCO-OFDM) satisfies these constraints by imposing Hermitian symmetry and adding an appropriate direct current (DC) bias.

Accurate PE and CE are essential for MIMO DCO-OFDM VL-ISAC operation. Practical environments contain only a few dominant targets and paths, producing sparsity across the reflection/path coefficients, delay, and Doppler domains \cite{10403776}. On-grid methods exploit this structure using predefined delay-Doppler (DD) grids but experience basis mismatch when the actual parameters lie between grid points. Off-grid formulations alleviate this limitation by estimating the sparse coefficients and grid deviations. Motivated by these considerations, this work develops a sparse PE and CE framework for a monostatic MIMO DCO-OFDM-aided VL-ISAC system that characterizes target reflections, clutter, Doppler effects, and the multipath communication channel. The relevant literature is reviewed next.

\vspace{-4mm}
% \begin{table*}[t]
% \centering
% \label{tab:comparison}
% \caption{Contrasting our contributions to the literature}%\vspace{-2mm}
% \resizebox{!}{!}{%
% \begin{tabular}{|l|c|c|c|c|c|c|c|c|c|}
% \hline
% \textbf{Features} 
% & \textbf{\cite{10901628}} 
% & \textbf{\cite{10697102}} 
% & \textbf{\cite{10960484}} 
% & \textbf{\cite{10900727}} 
% & \textbf{\cite{11267253}} 
% & \textbf{\cite{9738478}} 
% & \textbf{\cite{8537983}} 
% & \textbf{\cite{11101062}} 
% & \textbf{Proposed} \\

% \hline

% DCO-OFDM 
% & $\checkmark$ & $\checkmark$ &  &  &  &  &  &  & $\checkmark$ \\ \hline

% LoS/NLoS VL channel 
% &  & $\checkmark$ & $\checkmark$ & $\checkmark$ &  &  &  &  & $\checkmark$ \\ \hline

% ISAC framework 
% & $\checkmark$ & $\checkmark$ & $\checkmark$ & $\checkmark$ & $\checkmark$ &  &  & $\checkmark$ & $\checkmark$ \\ \hline

% BL-based estimation
% &  &  &  &  &  & $\checkmark$ & $\checkmark$ &  & $\checkmark$ \\ \hline

% BCRLB/CRLB analysis 
% &  &  &  &  &  &  &  & $\checkmark$ & $\checkmark$ \\ \hline

% $1$D sparse estimation 
% &  &  &  &  & $\checkmark$ & $\checkmark$ & $\checkmark$ &  & $\checkmark$ \\ \hline

% $2$D sparse estimation 
% &  &  &  &  &  & $\checkmark$ & $\checkmark$ &  & $\checkmark$ \\ \hline

% $1$D sparse off-grid estimation 
% &  &  &  &  &  & $\checkmark$ & $\checkmark$ &  & $\checkmark$ \\ \hline

% $2$D sparse off-grid estimation 
% &  &  &  &  &  & $\checkmark$ &  &  & $\checkmark$ \\ \hline

% Joint target and CE
% &  &  &  &  & $\checkmark$ &  &  &  & $\checkmark$ \\

% \hline
% \end{tabular}%
% }
% \vspace{-3mm}
% \end{table*}
\begin{table*}[t]
\centering
\caption{Boldly contrasting our contributions to the literature}
\label{tab:comparison}
\resizebox{!}{!}{%
\begin{tabular}{|l|c|c|c|c|c|c|c|c|c|c|}
\hline
\textbf{Features}
& \textbf{\cite{10960484}}
& \textbf{\cite{10901628}}
& \textbf{\cite{10697102}}
& \textbf{\cite{10900727}}
& \textbf{\cite{11101062}}
& \textbf{\cite{10817491}}
& \textbf{\cite{wen2025optical}}
& \textbf{\cite{11293749}}
& \textbf{\cite{saxena2025multiple}} & \textbf{Proposed} \\ \hline

VLC system
& $\checkmark$ & $\checkmark$ & $\checkmark$ & $\checkmark$
& $\checkmark$ & $\checkmark$ & $\checkmark$ & $\checkmark$
& $\checkmark$ & $\checkmark$ \\ \hline

VL-ISAC system
& $\checkmark$ & $\checkmark$ & $\checkmark$ & $\checkmark$
& $\checkmark$ & $\checkmark$ & $\checkmark$ & $\checkmark$
& & $\checkmark$ \\ \hline

Time delay/range estimation
& $\checkmark$ & & $\checkmark$ & $\checkmark$
& & & & & & $\checkmark$\\ \hline

Joint PE and CE
& $\checkmark$ & & $\checkmark$ & $\checkmark$
& & & & & & $\checkmark$\\ \hline

BCRLB/CRLB analysis
& & & & & $\checkmark$
& & & & $\checkmark$ & $\checkmark$\\ \hline

Unknown clutters
& & & & & & & $\checkmark$ & & & $\checkmark$\\ \hline

\textbf{MIMO-OOFDM}
& & & & & & & & & & $\checkmark$\\ \hline

\textbf{Doppler shift/velocity estimation}
& & & & & & & & & & $\checkmark$\\ \hline

\textbf{$1$D sparse C\&S representation}
& & & & & & & & & & $\checkmark$\\ \hline

\textbf{$2$D sparse C\&S representation}
& & & & & & & & & & $\checkmark$\\ \hline

\textbf{$1$D-OffGBL algorithm for PE and CE}
& & & & & & & & & & $\checkmark$\\ \hline

\textbf{$2$D-OffGBL algorithm for PE and CE}
& & & & & & & & & & $\checkmark$\\ \hline

\end{tabular}%
}
\end{table*}
%\vspace{-1mm}
\subsection{Literature review}
Most existing studies on ISAC have been conducted in the RF domain. These RF ISAC works mainly focus on designing unified waveforms that simultaneously support C\&S functionalities. In \cite{11267253}, a sensing-assisted CE framework was developed for bistatic OFDM-ISAC systems by utilizing sensing information to improve CE accuracy. The authors in \cite{10105893} investigated delay alignment modulation-based ISAC systems and proposed beamforming optimization techniques for simultaneous C\&S. A low-complexity sensing integration framework for communication-centric OFDM systems was presented in \cite{9727202}, where discrete Fourier transform spread OFDM and orthogonal time-frequency space signaling were utilized for flexible sensing operations. Furthermore, sparse vector coding-based waveform design was proposed in \cite{9695365} to achieve low sensing sidelobes and ultra-reliable communication performance. In these RF-ISAC systems, a common waveform is generally transmitted for communication, while the reflected echoes from surrounding targets are processed to extract sensing information. A comprehensive tutorial on MIMO-OFDM ISAC systems was presented in \cite{11328117}, where unified signal models and PE techniques across the spatial, delay, and Doppler domains were discussed.

Driven by spectrum scarcity and interference limitations in RF systems, VL-ISAC has emerged as a promising complement to RF-ISAC. The architectures, performance limits, and waveform design aspects of VL-ISAC were comprehensively reviewed in \cite{10574248}. In \cite{11101062}, the performance of O-ISAC was characterized from a capacity-distortion trade-off perspective, wherein practical maximum \textit{a posteriori} (MAP) and maximum likelihood estimators were developed for target distance estimation, together with the corresponding Bayesian Cramér-Rao bound (BCRB) analysis. The work in \cite{10817491} investigated LED-based VL-ISAC systems with optical beamforming for simultaneous C\&S, while \cite{wen2025optical} developed optimal beamforming techniques based on optical phased arrays for VL-ISAC systems. Furthermore, multiuser retroreflective VL-ISAC systems with joint waveform design and multiple-access strategies were investigated in \cite{11293749}.

%In \cite{11101062}, the fundamental performance limits and transceiver design aspects of VL-ISAC were investigated using Bayesian Cramér-Rao bound (BCRB) analysis, while \cite{10574248} presented generalized VL-ISAC architectures and waveform design strategies based on pulsed, constant-modulus, and multicarrier optical signals. Unlike RF-ISAC, VL-ISAC exploits the ultra-wide and license-free optical spectrum, enabling higher data rates, improved sensing precision, and reduced electromagnetic interference. In \cite{10960484}, an enhanced asymmetrically clipped direct current biased O-OFDM-based VL-ISAC framework with flexible resource allocation was proposed. 

Although OFDM-based ISAC has been extensively studied in RF systems, its direct extension to VL-ISAC is challenging due to the IM/DD constraint, which requires the transmitted optical signal to be real-valued and non-negative and may introduce nonlinear clipping distortion \cite{10901628,10960484}. Consequently, only a limited number of studies have considered O-OFDM-based VL-ISAC frameworks, with MIMO VL configurations remaining largely unexplored. In \cite{10901628}, adaptive resource allocation techniques were proposed for asymmetrically clipped DC-biased O-OFDM (ADO-OFDM)-based VL-ISAC systems. The authors in \cite{10697102} investigated DCO-OFDM-based free-space optical (FSO) ISAC systems with joint power allocation considering clipping noise and sensing accuracy. Flexible resource allocation for enhanced asymmetrically clipped DC-biased O-OFDM-based VL-ISAC systems was further studied in \cite{10960484}, while \cite{10622583} proposed power allocation schemes for OFDM-based FSO-ISAC systems under clipping distortion effects. However, these studies mainly focus on waveform design and resource allocation for VL and FSO links, while limited attention has been given to signal-level sensing PE and CE in MIMO DCO-OFDM-aided VL-ISAC systems.

In parallel, substantial research has been devoted to improving O-OFDM-based VLC systems. Since DCO-OFDM is widely adopted to satisfy the IM/DD constraint, accurate CE is essential for mitigating the effects of multipath propagation, ambient-light interference, and noise. Conventional estimators exhibit notable limitations: least squares (LS) is highly sensitive to noise, whereas linear minimum mean square error (LMMSE) entails increased computational complexity and requires prior channel statistics \cite{saxena2025multiple}. By exploiting the limited number of dominant propagation paths in practical VLC channels \cite{saxena2025multiple,10403776}, sparse recovery methods such as orthogonal matching pursuit (OMP) and FOCal
Underdetermined System Solver (FOCUSS) can utilize delay-domain sparsity to improve CE performance \cite{saxena2025multiple}. In \cite{saxena2023sparse}, sparse Bayesian learning (BL)-based sparse CE schemes were developed for DCO-OFDM and asymmetrically clipped O-OFDM (ACO-OFDM) VLC systems. Although these approaches generally outperform LS and LMMSE, their performance depends on the assumed sparsity level, dictionary resolution, and parameter selection. Moreover, grid mismatch, model inaccuracies, and ill-conditioned sensing matrices may impair convergence and estimation accuracy \cite{saxena2025multiple}. Off-grid formulations address these limitations by jointly estimating the sparse coefficients and grid deviations, as demonstrated through DD- or direction-of-arrival-domain refinement in \cite{9738478} and \cite{8537983}. Consequently, off-grid BL-based estimation provides a promising approach for enhancing robustness to dictionary mismatch.

Despite substantial advances in RF-ISAC, VL-ISAC, and O-OFDM-based VLC, the existing literature generally addresses communication, sensing, and parameter estimation independently. Most VL-ISAC studies focus primarily on waveform or resource allocation without adequately incorporating the multipath characteristics of VL channels. Conversely, O-OFDM-based VLC studies remain predominantly communication-oriented and do not consider integrated sensing capabilities. Consequently, a unified framework that incorporates O-OFDM signaling, MIMO transmission, multipath propagation, target reflections, Doppler effects, and clutter within a common VL-ISAC signal model remains largely unexplored. This research gap motivates the development of an integrated VL radar-communication framework that supports both CE and sensing PE. Table \ref{tab:comparison} compares the proposed framework with the relevant literature, while the principal contributions of this work are summarized as follows.

\vspace{-5mm}
\subsection{Contributions}
\begin{enumerate}
    \item A unified monostatic MIMO DCO-OFDM-aided VL-ISAC framework is formulated, wherein a shared intensity-modulated optical waveform enables simultaneous C\&S. The resulting model incorporates target reflections, clutter, delay and Doppler effects, and the communication channel while complying with the IM/DD constraints.

    \item For sensing PE, $1$D on-grid and off-grid sparse representations are formulated using a joint DD dictionary. Based on the off-grid model, a $1$D off-grid BL ($1$D-OffGBL) algorithm is developed to jointly estimate the sparse target coefficients and DD grid deviations, thereby mitigating basis mismatch and improving the estimation of continuous-valued sensing parameters.

    \item To reduce the computational complexity associated with the joint DD dictionary, $2$D on-grid and off-grid sparse representations are developed by separating the Doppler and delay estimation stages. A $2$D off-grid BL ($2$D-OffGBL) algorithm is subsequently proposed to sequentially estimate and refine the Doppler and delay parameters, providing a favorable trade-off between estimation accuracy and computational complexity.

    \item The proposed sensing framework is complemented by $1$D and $2$D sparse channel representations and corresponding Bayesian CE procedures. Systematic methods are also developed to recover the target range and velocity from the estimated coefficients of the respective $1$D and $2$D sensing models.

    \item Closed-form BCRBs are derived for sensing PE, CE, range, and velocity estimation, providing theoretical performance benchmarks. Simulation results demonstrate that the proposed algorithms outperform conventional sparse recovery schemes, with $1$D-OffGBL achieving superior estimation accuracy and $2$D-OffGBL offering reduced complexity with competitive performance.
\end{enumerate}
\vspace{-5mm}

\section{Radar and Communication VL System Models}
This section formulates the sensing and downlink communication models of the considered MIMO DCO-OFDM VL-ISAC system and describes the associated DCO-OFDM waveform construction.

\vspace{-4mm}
\subsection{MIMO DCO-OFDM VL-ISAC system model}

A monostatic MIMO architecture is considered, where an access point (AP) concurrently provides sensing and communication functionalities. The AP is equipped with $N_t$ LEDs and $N_r$ sensing photodetectors (PDs), whereas the optical receiver at the user equipment (UE) consists of $R$ PDs. The surrounding VL environment comprises $P$ reflecting objects involved in sensing, while $L$ dominant scatterers form the propagation channel from the AP to the UE. The available bandwidth $B$ is divided among $M$ mutually orthogonal subcarriers separated by $\Delta f$, yielding $B=M\Delta f$. A transmission frame containing $N$ DCO-OFDM symbols is employed to facilitate processing in the DD domain. The useful symbol interval is given by $T=1/\Delta f$, whereas the complete symbol duration is $T_s=T+T_{\mathrm{cp}}$. To prevent interference between consecutive symbols, $T_{\mathrm{cp}}$ is chosen to be longer than the maximum delay spread. Moreover, the condition $\Delta f>\nu_{\max}$ is imposed to suppress the inter-carrier interference caused by Doppler shifts \cite{10960484}.

For compatibility with IM/DD, the transmitted DCO-OFDM signal must be both real and non-negative \cite{saxena2025multiple}. Accordingly, the information bits are mapped onto complex constellation points, such as quadrature amplitude modulation (QAM) symbols. Let $x_t(n,m)$ represent the symbol assigned to the $m$-th subcarrier of the $n$-th DCO-OFDM symbol transmitted from LED $t$. A real-valued time-domain waveform is obtained by enforcing the Hermitian-symmetry conditions $x_t(n,i)=x_t^{*}(n,M-i)$, $i=1,\ldots,M/2-1$, with $x_t(n,0)=x_t(n,M/2)=0$, prior to applying the $M$-point IDFT \cite{saxena2023sparse}. The cyclic prefix (CP) is subsequently appended, followed by digital-to-analog conversion and low-pass filtering. A suitable DC offset is then introduced to shift the waveform into the non-negative region, and any remaining negative-valued samples are clipped \cite{saxena2023sparse}. The resulting unipolar electrical signal is supplied to the LED for optical transmission.
The received echo at the $r$th sensing PD is given by \cite{9540344}
\vspace{-3mm}
\begin{align}
y_{\mathrm{r}}(n,m)
= \sum_{t=1}^{N_t} x_t(n,m) h_{r,t}(n,m) + v_r(n,m),
\label{eq:received_signal}
\end{align}
where $v_r(n,m)\sim\mathcal{CN}(0,\sigma_v^2)$ denotes complex additive white Gaussian noise (AWGN). Assuming $P$ point targets in the sensing scene, the radar channel between the $t$th transmit LED and the $r$th receive PD is modeled as $h_{r,t}(n,m)=\sum_{p=1}^{P}\alpha_{r,t,p} e^{-j2\pi m\Delta f \tau_p} e^{j2\pi nT_s\nu_p}$, where $\alpha_{r,t,p}$, $\tau_p$, and $\nu_p$ denote the RCS, delay, and Doppler frequency of the $p$th target, respectively. Accordingly, the received target echo can be expressed as
\vspace{-3mm}
\begin{align}
y_{\mathrm{r}}(n,m)
&= \sum_{t=1}^{N_t}\sum_{p=1}^{P}
\alpha_{r,t,p} x_t(n,m)
e^{-j2\pi m\Delta f \tau_p}
e^{j2\pi nT_s\nu_p} \nonumber \\
&+ v_r(n,m).
\label{eq:target_echo}
\end{align}
Let $N_{\mathrm{cl}}$ denote the number of clutter scatterers, which are assumed to have zero Doppler shift. The clutter channel is given by $h^{\mathrm{cl}}_{r,t}(n,m)=\sum_{c=1}^{N_{\mathrm{cl}}}\alpha_{r,t,c} e^{-j2\pi m\Delta f \tau_{\mathrm{c}}}$, where $\alpha_{r,t,c}$ and $\tau_{\mathrm{c}}$ denote the reflection coefficient and delay of the $c$'th clutter component, respectively. Let $S=P+N_{\mathrm{cl}}$ denote the combined number of targets and clutter. The received echo can then be written in the unified form \cite{9540344}
\vspace{-2mm}
\begin{align}
y_{\mathrm{r}}(n,m)
&= \sum_{t=1}^{N_t}\sum_{s=1}^{S}
\alpha_{r,t,s} x_t(n,m)
e^{-j2\pi m\Delta f \tau_s}
e^{j2\pi nT_s\nu_s} \nonumber \\
&+ v_r(n,m),
\label{eq:unified_echo}
\end{align}
where $\nu_s=\nu_p$ for $s=1,\ldots,P$, while $\nu_s=0$ for $s=P+1,\ldots,S$.

For communication, assume that $L$ multipath components exist between the AP and the UE. The signal received at the $r$-th UE PD is given by \cite{9540344}
\vspace{-2mm}
\begin{align}
y_r^{\mathrm{c}}(n,m)
= \sum_{t=1}^{N_t} x_t(n,m) h^{\mathrm{c}}_{r,t}(n,m) + z_r(n,m),
\label{eq:comm_received}
\end{align}
where $z_r(n,m)\sim\mathcal{CN}(0,\sigma_z^2)$ denotes complex AWGN. %Since the UE employs a fully digital (FD) optical receiver, conventional linear combiners, such as zero-forcing (ZF) or minimum mean square error (MMSE) combiners, can be applied for symbol detection. 
The term $h^{\mathrm{c}}_{r,t}(n,m)$ denotes the downlink VLC MIMO DCO-OFDM channel frequency response (CFR) corresponding to the $m$th subcarrier and the $n$th DCO-OFDM symbol. It is modeled as
\vspace{-2mm}
\begin{align}
h^{\mathrm{c}}_{r,t}(n,m)
= \sum_{\ell=1}^{L}
\xi_{r,t,\ell}
e^{-j2\pi m\Delta f \tau_\ell}
e^{j2\pi nT_s\nu_\ell},
\label{eq:comm_channel}
\end{align}
where $\xi_{r,t,\ell}$, $\tau_\ell$, and $\nu_\ell$ denote the path gain, delay, and Doppler frequency of the $\ell$th multipath component, respectively \cite{9540344}. To account for the increased attenuation of delayed optical reflections, the path gains follow an exponential delay profile, given by 
\vspace{-2mm}
\begin{align}
\xi_{r,t,\ell}
=
\frac{
e^{-l_{r,t,\ell}\Delta\tau_\ell/\tau_{\mathrm{rms},r,t}}
}{
\displaystyle\sum_{\tilde{l}=0}^{L-1}
e^{-\tilde{l}\Delta\tau_\ell/\tau_{\mathrm{rms},r,t}}
},
\label{eq:exp_decay_mimo}
\end{align}
where $l_{r,t,\ell}\in\{0,\ldots,L-1\}$ identifies the delay bin associated with the $\ell$-th path, and $\tau_{\mathrm{rms},r,t}$ represents the root-mean-square (RMS) delay spread \cite{zhang2021evolution,saxena2023sparse,saxena2025multiple}. Based on the propagation geometry and reflective properties of the surrounding surfaces, $\tau_{\mathrm{rms},r,t}$ is generally selected within the interval $[0.5\Delta\tau_\ell,1.5\Delta\tau_\ell]$ \cite{zhang2021evolution,saxena2023sparse,saxena2025multiple}. Substituting \eqref{eq:comm_channel} into \eqref{eq:comm_received}, the received communication signal at the UE becomes 
\vspace{-2mm}
\begin{align}
y_r^{\mathrm{c}}(n,m)
&= \sum_{t=1}^{N_t}\sum_{\ell=1}^{L}
\xi_{r,t,\ell} x_t(n,m)
e^{-j2\pi m\Delta f \tau_\ell}
e^{j2\pi nT_s\nu_\ell} \nonumber \\
&+ z_r(n,m).
\label{eq:comm_received_expanded}
\end{align}
Since the C\&S channel contain only a few dominant propagation paths and targets within the considerably larger DD search space \cite{9540344}, both models exhibit inherent sparsity, motivating the $1$D sparse formulations for sensing PE and subsequent CE.
\vspace{-4mm}
\section{$1$D off-grid sparse modeling and BL-Based PE}
This section presents the problem formulation for $1$D sparse off-grid formulation, followed by the $1$D-OffGBL-based PE algorithm for the sensing model in the considered MIMO DCO-OFDM-aided VL-ISAC framework. 
\vspace{-4mm}
\subsection{$1$D on-grid sparse representation for sensing PE}
To obtain a sparse representation of the VL MIMO DCO-OFDM radar channel in the DD-domain, let $D_{\tau}$ and $D_{\nu}$ denote the delay and Doppler grid sizes, respectively. The corresponding grids are defined as $\mathcal{D}_{\tau}=\{\tau_i=i/(M\Delta f)\}_{i=0}^{D_{\tau}-1}$ and $\mathcal{D}_{\nu}=\{\nu_k=k/(NT_s)\}_{k=0}^{D_{\nu}-1}$. Let $\alpha_{r,t,i,k}$ denote the complex RCS coefficient associated with the $i$th delay grid point and the $k$th Doppler grid point for the $t$th transmit LED, and the $r$th receive PD. Assuming that the target parameters lie on the predefined DD grid, the radar channel between the $t$th transmit LED and the $r$th receive PD is expressed as 
\vspace{-2mm}
\begin{align}
h_{r,t}(n,m)
=
\sum_{i=0}^{D_{\tau}-1}
\sum_{k=0}^{D_{\nu}-1}
\alpha_{r,t,i,k}
e^{-j2\pi \frac{m}{M}i}
e^{j2\pi \frac{n}{N}k}.
\label{eq:dd_channel}
\end{align}
The received echo at the $r$th receive PD is given by $y_r(n,m)=\sum_{t=1}^{N_t}x_t(n,m)h_{r,t}(n,m)+v_r(n,m)$, where $v_r(n,m)$ denotes AWGN. Substituting \eqref{eq:dd_channel} into \eqref{eq:received_signal}, the received signal becomes
\vspace{-4mm}
\begin{align}
y_r(n,m)
=
\sum_{t=1}^{N_t}
\sum_{i=0}^{D_{\tau}-1}
\sum_{k=0}^{D_{\nu}-1}
\alpha_{r,t,i,k}
\beta_{i,k,t}(n,m)
+ v_r(n,m),
\label{eq:received_sparse_scalar}
\end{align}
where $\beta_{i,k,t}(n,m)=x_t(n,m)e^{-j2\pi \frac{m}{M}i}e^{j2\pi \frac{n}{N}k}$. By stacking the received samples over all $N$ DCO-OFDM symbols and $M$ subcarriers, the observation vector is defined as $\mathbf{y}_r=[y_r(0,0),y_r(0,1),\ldots,y_r(N-1,M-1)]^T \in \mathbb{C}^{MN\times 1}$. For the $t$th transmit LED, the sensing dictionary is defined as $\boldsymbol{\Phi}_t=[\boldsymbol{\beta}_{0,0,t},\boldsymbol{\beta}_{0,1,t},\ldots,\boldsymbol{\beta}_{D_{\tau}-1,D_{\nu}-1,t}]\in\mathbb{C}^{MN\times D_{\tau}D_{\nu}}$, where $\boldsymbol{\beta}_{i,k,t}=[\beta_{i,k,t}(0,0),\ldots,\beta_{i,k,t}(N-1,M-1)]^T$. Similarly, the sparse coefficient vector is $\boldsymbol{\alpha}_{r,t}=[\alpha_{r,t,0,0},\alpha_{r,t,0,1},\ldots,\alpha_{r,t,D_{\tau}-1,D_{\nu}-1}]^T\in\mathbb{C}^{D_{\tau}D_{\nu}\times 1}$. Thus, the received signal can be written as
\vspace{-3mm}
\begin{align}
\mathbf{y}_r
=
\sum_{t=1}^{N_t}
\boldsymbol{\Phi}_t
\boldsymbol{\alpha}_{r,t}
+
\mathbf{v}_r.
\label{eq:single_rx_sparse}
\end{align}
Stacking the dictionaries and coefficient vectors over all transmit LEDs yields $\boldsymbol{\Phi}_{\mathrm{MIMO}}=[\boldsymbol{\Phi}_1,\boldsymbol{\Phi}_2,\ldots,\boldsymbol{\Phi}_{N_t}]\in\mathbb{C}^{MN\times D_{\tau}D_{\nu}N_t}$ and $\boldsymbol{\alpha}_r=[\boldsymbol{\alpha}_{r,1}^T,\boldsymbol{\alpha}_{r,2}^T,\ldots,\boldsymbol{\alpha}_{r,N_t}^T]^T$. Hence, the MIMO sparse observation model for the $r$th receive PD is $\mathbf{y}_r
=
\boldsymbol{\Phi}_{\mathrm{MIMO}}
\boldsymbol{\alpha}_r
+
\mathbf{v}_r.$
% \begin{align}
% \mathbf{y}_r
% =
% \boldsymbol{\Phi}_{\mathrm{MIMO}}
% \boldsymbol{\alpha}_r
% +
% \mathbf{v}_r.
% \label{eq:mimo_sparse}
% \end{align}
Finally, stacking the observations from all $N_r$ receive PDs gives
\vspace{-2mm}
\begin{align}
\mathbf{Y}
=
\boldsymbol{\Phi}_{\mathrm{MIMO}}
\mathbf{A}
+
\mathbf{V},
\label{eq:mimo_sparse_multi_rx}
\end{align}
where $\mathbf{Y}=[\mathbf{y}_1,\ldots,\mathbf{y}_{N_r}]$, $\mathbf{A}=[\boldsymbol{\alpha}_1,\ldots,\boldsymbol{\alpha}_{N_r}]$, and $\mathbf{V}=[\mathbf{v}_1,\ldots,\mathbf{v}_{N_r}]$. The coefficient vectors $\boldsymbol{\alpha}_r$ are sparse as the number of dominant target and clutter components $S$ is much smaller than the DD dictionary size, i.e., $S\ll D_{\tau}D_{\nu}$ \cite{9540344}.

\vspace{-4mm}
\subsection{$1$D off-grid sparse representation for sensing PE}
In this subsection, based on the DD-domain sensing input-output relation, the target PE problem is formulated as a $1$D off-grid sparse signal representation (SSR) problem.
%\subsubsection{Sensing PE problem formulation}
For off-grid sensing PE, the received echo at the $r$th receive PD is expressed as \cite{9738478}
\vspace{-2mm}
\begin{align}
\mathbf{y}_r
=
\boldsymbol{\Phi}_{\mathrm{MIMO}}(\boldsymbol{\tau},\boldsymbol{\nu})
\boldsymbol{\alpha}_r
+
\mathbf{v}_r.
\label{eq:offgrid_basic}
\end{align}
%where $\mathbf{y}_r\in\mathbb{C}^{MN\times 1}$ denotes the vectorized received signal, $\mathbf{v}_r\sim\mathcal{CN}(\mathbf{0},\sigma^2\mathbf{I}_{MN})$ is the AWGN, and $\boldsymbol{\alpha}_r=[\boldsymbol{\alpha}_{r,1}^T,\ldots,\boldsymbol{\alpha}_{r,N_t}^T]^T$ contains the complex scattering coefficients corresponding to the $r$th receive PD. 
The vectors $\boldsymbol{\tau}=[\tau_1,\tau_2,\ldots,\tau_{D_\tau D_\nu}]^T$ and $\boldsymbol{\nu}=[\nu_1,\nu_2,\ldots,\nu_{D_\tau D_\nu}]^T$ collect the continuous-valued delay and Doppler grid points, respectively. For the $t$th transmit LED, the measurement matrix is defined as $\boldsymbol{\Phi}_t(\boldsymbol{\tau},\boldsymbol{\nu})=[\boldsymbol{\phi}_t(\tau_1,\nu_1),\ldots,\boldsymbol{\phi}_t(\tau_{D_\tau D_\nu},\nu_{D_\tau D_\nu})]\in\mathbb{C}^{MN\times D_\tau D_\nu}$, where the $(n,m)$th entry of the steering vector is given by $\phi_t(\tau_i,\nu_i)[n,m]=x_t(n,m)e^{-j2\pi m\Delta f\tau_i}e^{j2\pi nT_s\nu_i}$.

%\subsubsection{First-order linear approximation}
Following off-grid compressed sensing and using first-order linear approximation, a virtual sampling grid is introduced in the DD-domain. Let the delay and Doppler grid points be $\bar{\tau}_i=ir_\tau$ and $\bar{\nu}_j=jr_\nu$, where $r_\tau=2\tau_{\max}/D_\tau$ and $r_\nu=2\nu_{\max}/D_\nu$ denote the virtual delay and Doppler resolutions, respectively \cite{8537983}. The true parameters are assumed to lie close to their nearest grid points, i.e., $\tau_i=\bar{\tau}_i+\epsilon_{\tau_i}$ and $\nu_j=\bar{\nu}_j+\epsilon_{\nu_j}$, where $\epsilon_{\tau_i}\in[-r_\tau/2,r_\tau/2]$ and $\epsilon_{\nu_j}\in[-r_\nu/2,r_\nu/2]$ denote the corresponding off-grid deviations. Using a first-order Taylor expansion around $(\bar{\tau}_i,\bar{\nu}_j)$, the steering vector can be approximated as $\boldsymbol{\phi}_t(\tau_i,\nu_j)\approx \boldsymbol{\phi}_t(\bar{\tau}_i,\bar{\nu}_j)+\epsilon_{\tau_i}\frac{\partial \boldsymbol{\phi}_t}{\partial \tau}+\epsilon_{\nu_j}\frac{\partial \boldsymbol{\phi}_t}{\partial \nu}$, where higher-order terms are neglected. The corresponding element-wise derivatives are
\vspace{-1mm}
\begin{align}
\frac{\partial \phi_t}{\partial \tau}
=
-j2\pi m\Delta f\, \phi_t(\bar{\tau}_i,\bar{\nu}_j),
~~
\frac{\partial \phi_t}{\partial \nu}
=
j2\pi nT_s\, \phi_t(\bar{\tau}_i,\bar{\nu}_j).
\label{eq:offgrid_derivatives}
\end{align}
% \vspace{-2mm}
% \subsubsection{Off-grid SSR model}
Substituting the first-order approximation into \eqref{eq:offgrid_basic}, the received signal can be given as \cite{9738478,8537983}
\vspace{-2mm}
\begin{align}
\mathbf{y}_r
&=
\left(
\boldsymbol{\Phi}_{\mathrm{MIMO}}
+
\boldsymbol{\Phi}_{\mathrm{MIMO},\tau}\mathrm{diag}(\boldsymbol{\epsilon}_{\tau})
+
\boldsymbol{\Phi}_{\mathrm{MIMO},\nu}\mathrm{diag}(\boldsymbol{\epsilon}_{\nu})
\right)
\boldsymbol{\alpha}_r \nonumber \\
&+
\mathbf{v}_r,
\label{eq:offgrid_expanded}
\end{align}
where $\boldsymbol{\Phi}_{\mathrm{MIMO}}$ is the on-grid dictionary, while $\boldsymbol{\Phi}_{\mathrm{MIMO},\tau}$ and $\boldsymbol{\Phi}_{\mathrm{MIMO},\nu}$ denote the derivative dictionaries with respect to delay and Doppler, respectively. Defining the effective off-grid dictionary as $\tilde{\boldsymbol{\Phi}}_{\mathrm{MIMO}}=\boldsymbol{\Phi}_{\mathrm{MIMO}}+\boldsymbol{\Phi}_{\mathrm{MIMO},\tau}\mathrm{diag}(\boldsymbol{\epsilon}_{\tau})+\boldsymbol{\Phi}_{\mathrm{MIMO},\nu}\mathrm{diag}(\boldsymbol{\epsilon}_{\nu})$, the compact observation model becomes $\mathbf{y}_r
=
\tilde{\boldsymbol{\Phi}}_{\mathrm{MIMO}}
\boldsymbol{\alpha}_r
+
\mathbf{v}_r.$
% \begin{align}
% \mathbf{y}_r
% =
% \tilde{\boldsymbol{\Phi}}_{\mathrm{MIMO}}
% \boldsymbol{\alpha}_r
% +
% \mathbf{v}_r.
% \label{eq:offgrid_compact}
% \end{align}
Stacking the received signals from all $N_r$ receive PDs yields
\vspace{-2mm}
\begin{align}
\mathbf{Y}
=
\tilde{\boldsymbol{\Phi}}_{\mathrm{MIMO}}
\mathbf{A}
+
\mathbf{V},
\label{eq:offgrid_multi_rx}
\end{align}
where $\mathbf{Y}\in\mathbb{C}^{MN\times N_r}$, $\mathbf{A}$ is jointly sparse, and all its columns share a common support. The on-grid indices provide coarse target locations, while the off-grid deviation vectors $\boldsymbol{\epsilon}_{\tau}$ and $\boldsymbol{\epsilon}_{\nu}$ refine the delay and Doppler estimates beyond the grid resolution. When $\boldsymbol{\epsilon}_{\tau}=\boldsymbol{\epsilon}_{\nu}=\mathbf{0}$, the proposed formulation reduces to the conventional on-grid sparse sensing PE model.

\vspace{-3mm}
\subsection{$1$D off-grid BL ($1$D-OffGBL)-based target PE}
BL is employed to exploit the sparsity of the RCS coefficient matrix $\mathbf{A}$ and estimate its non-zero entries. Let $\boldsymbol{\alpha}_g$ denote the $g$th row of $\mathbf{A}$, where $g=0,1,\ldots,G-1$ and $G=D_\tau D_\nu N_t$. In the BL framework, each row $\boldsymbol{\alpha}_g$ is assigned a parameterized complex Gaussian prior as \cite{dosso2008bayesian}
\vspace{-2mm}
\begin{align}
p(\boldsymbol{\alpha}_g;z_g)
=
\mathcal{CN}\left(\mathbf{0}_{N_r\times 1},z_g^{-1}\mathbf{I}_{N_r}\right)
=
\prod_{r=1}^{N_r}
\frac{z_g}{\pi}
\exp\left(-z_g|\alpha_{g,r}|^2\right),
\label{eq:BL_prior_row}
\end{align}
where $z_g$ is the precision hyperparameter associated with $\boldsymbol{\alpha}_g$. The prior distribution of $\mathbf{A}$ is therefore given by $p(\mathbf{A};\mathbf{z})=\prod_{g=0}^{G-1}p(\boldsymbol{\alpha}_g;z_g)$, where $\mathbf{z}=[z_0,z_1,\ldots,z_{G-1}]^T$. As $z_g\rightarrow\infty$, the corresponding row $\boldsymbol{\alpha}_g$ tends to $\mathbf{0}_{N_r\times 1}$, thereby enforcing row sparsity in $\mathbf{A}$. To avoid over-parameterization and promote sparsity, a Gamma prior is imposed on the hyperparameter vector $\mathbf{z}$ as \cite{1315936}
\vspace{-2mm}
\begin{align}
p(\mathbf{z})
=
\prod_{g=0}^{G-1}
\mathrm{Gamma}(z_g|a+1,b)
=
\prod_{g=0}^{G-1}
\frac{b^a z_g^a e^{-bz_g}}{\Gamma(a+1)},
\label{eq:gamma_prior}
\end{align}
where $\Gamma(a)=\int_{0}^{\infty}t^{a-1}e^{-t}dt$ denotes the Gamma function. To obtain a non-informative prior, the hyperparameters are set to small values, e.g., $a=b=10^{-4}$. The off-grid deviation vectors $\boldsymbol{\epsilon}_{\tau}$ and $\boldsymbol{\epsilon}_{\nu}$ are also assigned non-informative uniform priors. Let $\boldsymbol{\Omega}_{\alpha}=\{\mathbf{A},\mathbf{z},\boldsymbol{\epsilon}_{\tau},\boldsymbol{\epsilon}_{\nu}\}$ denote the set of hidden variables \cite{7547360}. The MAP estimation problem is given as
%\vspace{-2mm}
\begin{align}
\widehat{\boldsymbol{\Omega}}_{\alpha}
=
\arg\max_{\boldsymbol{\Omega}_{\alpha}}
\ln p(\boldsymbol{\Omega}_{\alpha}|\mathbf{Y})
\equiv
\arg\max_{\boldsymbol{\Omega}_{\alpha}}
\ln p(\mathbf{Y},\boldsymbol{\Omega}_{\alpha}).
\label{eq:map_problem}
\end{align}
The posterior distribution is given by
%\vspace{-2mm}
\begin{align}
p(\boldsymbol{\Omega}_{\alpha}|\mathbf{Y})
=
\frac{
p(\mathbf{Y}|\mathbf{z},\boldsymbol{\epsilon}_{\tau},\boldsymbol{\epsilon}_{\nu})p(\mathbf{z})
}{
p(\mathbf{Y})
},
\label{eq:posterior_BL}
\end{align}
where $p(\mathbf{Y})$ is the Bayesian evidence. The likelihood is modeled as $p(\mathbf{Y}|\mathbf{A},\mathbf{z},\boldsymbol{\epsilon}_{\tau},\boldsymbol{\epsilon}_{\nu})=\mathcal{CN}(\mathbf{Y}|\tilde{\boldsymbol{\Phi}}_{\mathrm{MIMO}}\mathbf{A},\sigma^2\mathbf{I})$. Since the multidimensional integration required for computing \eqref{eq:posterior_BL} is analytically intractable, a block majorization-minimization approach is adopted. At a fixed point $(\dot{\mathbf{z}},\dot{\boldsymbol{\epsilon}}_{\tau},\dot{\boldsymbol{\epsilon}}_{\nu})$, a surrogate lower-bound function for $\ln p(\mathbf{Y},\mathbf{z},\boldsymbol{\epsilon}_{\tau},\boldsymbol{\epsilon}_{\nu})$ is constructed as
\vspace{-2mm}
\begin{align}
\mathcal{L}
&\left(
\mathbf{z},\boldsymbol{\epsilon}_{\tau},\boldsymbol{\epsilon}_{\nu}
\middle|
\dot{\mathbf{z}},\dot{\boldsymbol{\epsilon}}_{\tau},\dot{\boldsymbol{\epsilon}}_{\nu}
\right) \notag \\
&=
\int
p(\mathbf{A}|\mathbf{Y},\dot{\mathbf{z}},
\dot{\boldsymbol{\epsilon}}_{\tau},\dot{\boldsymbol{\epsilon}}_{\nu})
\ln
\frac{
p(\mathbf{A},\mathbf{Y},\mathbf{z},\boldsymbol{\epsilon}_{\tau},\boldsymbol{\epsilon}_{\nu})
}{
p(\mathbf{A}|\mathbf{Y},\dot{\mathbf{z}},
\dot{\boldsymbol{\epsilon}}_{\tau},\dot{\boldsymbol{\epsilon}}_{\nu})
}
d\mathbf{A}.
\label{eq:surrogate_function}
\end{align}
The parameters are then updated sequentially as
\vspace{-2mm}
\begin{align}
\mathbf{z}^{(k+1)}
&=
\arg\max_{\mathbf{z}}
\mathcal{L}\left(
\mathbf{z},\boldsymbol{\epsilon}_{\tau}^{(k)},\boldsymbol{\epsilon}_{\nu}^{(k)}
\middle|
\mathbf{z}^{(k)},\boldsymbol{\epsilon}_{\tau}^{(k)},\boldsymbol{\epsilon}_{\nu}^{(k)}
\right),  \\
\boldsymbol{\epsilon}_{\tau}^{(k+1)}
&=
\arg\max_{\boldsymbol{\epsilon}_{\tau}}
\mathcal{L}\left(
\mathbf{z}^{(k+1)},\boldsymbol{\epsilon}_{\tau},\boldsymbol{\epsilon}_{\nu}^{(k)}
\middle|
\mathbf{z}^{(k+1)},\boldsymbol{\epsilon}_{\tau}^{(k)},\boldsymbol{\epsilon}_{\nu}^{(k)}
\right),  \\
\boldsymbol{\epsilon}_{\nu}^{(k+1)}
&=
\arg\max_{\boldsymbol{\epsilon}_{\nu}}
\mathcal{L}\left(
\mathbf{z}^{(k+1)},\boldsymbol{\epsilon}_{\tau}^{(k+1)},\boldsymbol{\epsilon}_{\nu}
\middle|
\mathbf{z}^{(k+1)},\boldsymbol{\epsilon}_{\tau}^{(k+1)},\boldsymbol{\epsilon}_{\nu}^{(k)}
\right).
\label{eq:block_mm_updates}
\end{align}
This block-wise update strategy iteratively refines the sparsity-controlling hyperparameters and the off-grid DD deviations, thereby improving the accuracy of the estimated RCS matrix $\mathbf{A}$.
The posterior distribution of the $r$th column of the RCS coefficient matrix $\mathbf{A}$ is obtained using the BL framework. For the off-grid sparse sensing model, the $r$th measurement vector is given by $\mathbf{y}_r=\tilde{\boldsymbol{\Phi}}_{\mathrm{MIMO}}\boldsymbol{\alpha}_r+\mathbf{v}_r$, where $\tilde{\boldsymbol{\Phi}}_{\mathrm{MIMO}}$ denotes the effective off-grid dictionary and $\mathbf{v}_r\sim\mathcal{CN}(\mathbf{0},\sigma^2\mathbf{I})$. Assuming complex Gaussian noise, the likelihood is given by $p(\mathbf{y}_r|\boldsymbol{\alpha}_r)=\mathcal{CN}(\tilde{\boldsymbol{\Phi}}_{\mathrm{MIMO}}\boldsymbol{\alpha}_r,\sigma^2\mathbf{I})$. Further, using the Gaussian prior assigned to $\boldsymbol{\alpha}_r$, Bayes' rule yields $p(\boldsymbol{\alpha}_r|\mathbf{y}_r)\propto p(\mathbf{y}_r|\boldsymbol{\alpha}_r)p(\boldsymbol{\alpha}_r|\mathbf{Z})$, where $\mathbf{Z}=\mathrm{diag}(z_0,z_1,\ldots,z_{G-1})$ is the precision matrix. By completing the square in the exponent, the posterior distribution is obtained as $q(\boldsymbol{\alpha}_r)
=
\mathcal{CN}\left(
\boldsymbol{\mu}_r^{(k+1)},
\boldsymbol{\Sigma}^{(k+1)}
\right),$
where the posterior covariance and mean are given by \cite{1315936}
\vspace{-2mm}
\begin{align}
\boldsymbol{\Sigma}^{(k+1)}
&=
\left(
\frac{1}{\sigma^{2}}
\tilde{\boldsymbol{\Phi}}_{\mathrm{MIMO}}^{H}
\tilde{\boldsymbol{\Phi}}_{\mathrm{MIMO}}
+
\mathbf{Z}^{(k+1)}
\right)^{-1},  \\
\boldsymbol{\mu}_r^{(k+1)}
&=
\frac{1}{\sigma^{2}}
\boldsymbol{\Sigma}^{(k+1)}
\tilde{\boldsymbol{\Phi}}_{\mathrm{MIMO}}^{H}
\mathbf{y}_r .
\label{eq:posterior_mean_cov}
\end{align}
Stacking the posterior means corresponding to all $N_r$ receive PDs, we obtain
\vspace{-2mm}
\begin{align}
\mathbf{U}^{(k+1)}
=
\left[
\boldsymbol{\mu}_{1}^{(k+1)},
\ldots,
\boldsymbol{\mu}_{N_r}^{(k+1)}
\right]
=
\frac{1}{\sigma^{2}}
\boldsymbol{\Sigma}^{(k+1)}
\tilde{\boldsymbol{\Phi}}_{\mathrm{MIMO}}^{H}
\mathbf{Y}.
\label{eq:posterior_mean_matrix}
\end{align}
The sparsity-controlling hyperparameters and the off-grid deviations are then updated iteratively. The precision hyperparameter associated with the $g$th row of $\mathbf{A}$ is updated as \cite{7547360}
\vspace{-2mm}
\begin{align}
z_g^{(k+1)}
=
\frac{a+N_r}
{b+\sum_{r=1}^{N_r}
\left[
\boldsymbol{\mu}_{r}^{(k+1)}
\boldsymbol{\mu}_{r}^{(k+1)H}
+
\boldsymbol{\Sigma}^{(k+1)}
\right]_{g,g}},
\label{eq:z_update}
\end{align}
where $0\le g\le G-1$. The off-grid delay and Doppler deviation vectors are updated as
\vspace{-2mm}
\begin{align}
\boldsymbol{\epsilon}_{\tau}^{(k+1)}
&=
\left(
\Re\left\{
\boldsymbol{\Phi}_{\mathrm{MIMO},\tau}^{H}
\boldsymbol{\Phi}_{\mathrm{MIMO},\tau}
\odot
\left(
\mathbf{U}\mathbf{U}^{H}
+
N_r\boldsymbol{\Sigma}
\right)
\right\}
\right)^{\dagger}
\mathbf{p}_{\tau},
\label{eq:offgrid_update_eps_tau}
 \\ 
\boldsymbol{\epsilon}_{\nu}^{(k+1)}
&=
\left(
\Re\left\{
\boldsymbol{\Phi}_{\mathrm{MIMO},\nu}^{H}
\boldsymbol{\Phi}_{\mathrm{MIMO},\nu}
\odot
\left(
\mathbf{U}\mathbf{U}^{H}
+
N_r\boldsymbol{\Sigma}
\right)
\right\}
\right)^{\dagger}
\mathbf{p}_{\nu},
\label{eq:offgrid_updates}
\end{align}
where $(\cdot)^{\dagger}$ denotes the Moore-Penrose pseudo-inverse, $\odot$ represents the Hadamard product, and $\Re\{\cdot\}$ extracts the real part. The auxiliary vectors $\mathbf{p}_{\tau}$ and $\mathbf{p}_{\nu}$ are defined as
\vspace{-2mm}
\begin{align}
\mathbf{p}_{\tau}
&=
\Re\left\{
\sum_{r=1}^{N_r}
\operatorname{diag}(\boldsymbol{\mu}_{r})
\boldsymbol{\Phi}_{\mathrm{MIMO},\tau}^{H}
\left(
\mathbf{y}_{r}
-
\boldsymbol{\Psi}_{\mathrm{MIMO},\tau}
\boldsymbol{\mu}_{r}
\right)
\right\}
\notag \\
&\quad
-
N_r
\Re\left\{
\operatorname{diag}
\left(
\boldsymbol{\Phi}_{\mathrm{MIMO},\tau}^{H}
\boldsymbol{\Psi}_{\mathrm{MIMO},\tau}
\boldsymbol{\Sigma}
\right)
\right\},
\label{eq:p_tau}
\\
\mathbf{p}_{\nu}
&=
\Re\left\{
\sum_{r=1}^{N_r}
\operatorname{diag}(\boldsymbol{\mu}_{r})
\boldsymbol{\Phi}_{\mathrm{MIMO},\nu}^{H}
\left(
\mathbf{y}_{r}
-
\boldsymbol{\Psi}_{\mathrm{MIMO},\nu}
\boldsymbol{\mu}_{r}
\right)
\right\}
\notag \\
&\quad
-
N_r
\Re\left\{
\operatorname{diag}
\left(
\boldsymbol{\Phi}_{\mathrm{MIMO},\nu}^{H}
\boldsymbol{\Psi}_{\mathrm{MIMO},\nu}
\boldsymbol{\Sigma}
\right)
\right\}.
\label{eq:p_nu}
\end{align}
Here, $\boldsymbol{\Psi}_{\mathrm{MIMO},\tau}
=
\boldsymbol{\Phi}(\boldsymbol{\tau},\boldsymbol{\nu})
+
\boldsymbol{\Phi}_{\mathrm{MIMO},\nu}
\operatorname{diag}(\boldsymbol{\epsilon}_{\nu})$ and
$\boldsymbol{\Psi}_{\mathrm{MIMO},\nu}
=
\boldsymbol{\Phi}(\boldsymbol{\tau},\boldsymbol{\nu})
+
\boldsymbol{\Phi}_{\mathrm{MIMO},\tau}
\operatorname{diag}(\boldsymbol{\epsilon}_{\tau})$.
After estimating the off-grid deviations, the DD grids are refined as \cite{9738478}
\vspace{-2mm}
\begin{align}
\boldsymbol{\tau}^{(k+1)}
=
\boldsymbol{\tau}^{(k)}
+
\boldsymbol{\epsilon}_{\tau}^{(k+1)},
\qquad
\boldsymbol{\nu}^{(k+1)}
=
\boldsymbol{\nu}^{(k)}
+
\boldsymbol{\epsilon}_{\nu}^{(k+1)}.
\label{eq:grid_update}
\end{align}
The above steps are repeated until convergence. Since the hyperparameters determine the row sparsity of $\widehat{\mathbf{A}}$, the final estimate can be further refined by pruning rows associated with large precision values \cite{9738478}. Specifically, for a chosen threshold $z_{\mathrm{th}}$, the pruned RCS matrix is obtained as \cite{8718676}
\vspace{-2mm}
\begin{align}
\widehat{\mathbf{A}}(g,:)
=
\begin{cases}
\widehat{\mathbf{A}}(g,:), & z_g \le z_{\mathrm{th}},\\
\mathbf{0}, & z_g > z_{\mathrm{th}},
\end{cases}
\label{eq:pruned_A}
\end{align}
where $0\le g\le G-1$. The detailed derivations of the update rules in \eqref{eq:z_update} to \eqref{eq:grid_update} are provided in the Appendix. Although $1$D-OffGBL provides accurate joint DD estimation, its high-dimensional dictionary entails substantial complexity, motivating the sequential $2$D-OffGBL formulation as a favorable performance-complexity trade-off.

\begin{algorithm}[t]
\caption{$1$D-OffGBL-assisted target PE}
\label{sfblms_algo}
\small
\DontPrintSemicolon

\KwIn{Received observation matrix $\mathbf{Y}\in\mathbb{C}^{MN\times N_r}$, measurement matrix $\tilde{\boldsymbol{\Phi}}_{\mathrm{MIMO}}\in\mathbb{C}^{MN\times \mathcal{D}_\tau \mathcal{D}_\nu N_t}$, noise precision matrix $\sigma^{-2}\mathbf{I}_{\mathcal{D}_\tau \mathcal{D}_\nu N_t}$, and stopping parameters $\epsilon$ and $t_{\max}$.}

\KwOut{Estimated $\widehat{\boldsymbol{\mathrm{A}}}$, $\tilde{\boldsymbol{\tau}}$, $\tilde{\boldsymbol{\nu}}$.}

\textbf{Initialization:} Set $\widehat{z}_j^{(0)}=1$ for all $1\leq j\leq \mathcal{D}_\tau \mathcal{D}_\nu N_t$, and hence $\widehat{\mathbf{Z}}^{(0)}=\mathbf{I}_{\mathcal{D}_\tau \mathcal{D}_\nu N_t}$
Set $\mathbf{z}^{(-1)}=\mathbf{0}$, $\widehat{\mathbf{Z}}^{(-1)}=\mathbf{0}$, and $t=0$.
Initialize $\boldsymbol{\epsilon}_{\tau}^{(0)}$ and $\boldsymbol{\epsilon}_{\nu}^{(0)}$ as given in Section III-B
%\tcp{Apply row-wise separately, if required}

\While{$\left\lVert \boldsymbol{z}^{(t)}-\boldsymbol{z}^{(t-1)}\right\rVert_2 > \epsilon$ \text{and} $t<t_{\max}$}{
    
    \textbf{E-step:} Compute the \textit{a posteriori} covariance and mean
    \vspace{-2mm}
    \[
    \begin{aligned}
    \boldsymbol{\Sigma}^{(t+1)}
    &=
    \left(
    \frac{1}{\sigma^{2}}
    \tilde{\boldsymbol{\Phi}}_{\mathrm{MIMO}}^{H}
    \tilde{\boldsymbol{\Phi}}_{\mathrm{MIMO}}
    +
    \mathbf{Z}^{(t)}
    \right)^{-1},\\
    \mathbf{U}^{(t+1)}
    &=
    \left[
    \boldsymbol{\mu}_{1}^{(t+1)},
    \ldots,
    \boldsymbol{\mu}_{N_r}^{(t+1)}
    \right]
    =
    \frac{1}{\sigma^{2}}
    \boldsymbol{\Sigma}^{(t+1)}
    \tilde{\boldsymbol{\Phi}}_{\mathrm{MIMO}}^{H}
    \mathbf{Y}.
    \end{aligned}
    \]

    \textbf{M-step:}

    \textbf{(i) Update hyperparameter vector $\mathbf{z}^{(t+1)}$:}\\
    \For{$g=1,2,\ldots,\mathcal{D}_\tau \mathcal{D}_\nu N_t$ \vspace{-1mm}}{
    
        \[
        \vspace{-2mm}
        z_g^{(t+1)}
        =
        \frac{a+N_r}
        {\,b+\sum_{r=1}^{N_r}
        \Upsilon_r\left(
        \boldsymbol{\epsilon}_{\tau}^{(t)},
        \boldsymbol{\epsilon}_{\nu}^{(t)}
        \right)}
        \]
        Update $\Upsilon_r$ as: \vspace{-3mm}
    \[ 
    \begin{aligned}
    \mathbf{z}^{(t+1)}
    &=
    \left[
    z_1^{(t+1)},\ldots,z_{\mathcal{D}_\tau \mathcal{D}_\nu N_t}^{(t+1)}
    \right]^T,\\
    \mathbf{Z}^{(t+1)}
    &=
    \operatorname{diag}\left(\mathbf{z}^{(t+1)}\right).
    \end{aligned}
    \] }
    \textbf{(ii) Update off-grid parameters:}
    Update $\mathbf{p}_{\tau}$ and $\mathbf{p}_{\nu}$ according to \eqref{eq:p_tau} and \eqref{eq:p_nu},
    Update $\boldsymbol{\epsilon}_{\tau}^{(t+1)}$ and $\boldsymbol{\epsilon}_{\nu}^{(t+1)}$ according to \eqref{eq:offgrid_update_eps_tau} and \eqref{eq:offgrid_updates}

    Set $t\leftarrow t+1$
}

\textbf{return:} $\widehat{\boldsymbol{\mathrm{A}}}_{\mathrm{off}} = \mathbf{U}^{(t)},
\tilde{\boldsymbol{\tau}}^{(t)} = \tilde{\boldsymbol{\tau}}^{(0)} + \boldsymbol{\epsilon}_{\tau}^{(t)},
\tilde{\boldsymbol{\nu}}^{(t)} = \tilde{\boldsymbol{\nu}}^{(0)} + \boldsymbol{\epsilon}_{\nu}^{(t)}.$
\label{1d_offgrid_algorithm}
\end{algorithm}

\vspace{-4mm}
\section{$2$D Off-Grid sparse modeling and BL-based PE}
This section presents the $2$D on-grid sparse representation problem formulation, followed by the $2$D off-grid BL ($2$D-OffGBL)-based PE algorithm for the considered MIMO DCO-OFDM-aided VL-based sensing framework. 
\vspace{-4mm}
\subsection{$2$D on-grid sparse representation for sensing PE}
When the pilot sequence is separable across the Doppler and delay dimensions, the transmitted pilot matrix for the $t$th transmit LED can be expressed as $\mathbf{X}_t=\mathbf{x}_{\nu,t}\mathbf{x}_{\tau,t}^{H}\in\mathbb{C}^{N\times M}$, where $\mathbf{x}_{\nu,t}\in\mathbb{C}^{N\times 1}$ and $\mathbf{x}_{\tau,t}\in\mathbb{C}^{M\times 1}$ collect the pilot symbols across the time and frequency domains, respectively \cite{11039779}. Under this separable structure, the DD-domain received signal at the $r$th receive PD can be written as
\vspace{-2mm}
\begin{align}
y_r(n,m)
&=
\sum_{t=1}^{N_t}
\sum_{i=0}^{D_{\tau}-1}
\sum_{k=0}^{D_{\nu}-1}
\alpha_{r,t,k,i}
x_{\nu,t}[n]x_{\tau,t}^{*}[m]
e^{-j2\pi \frac{m}{M}i}
e^{j2\pi \frac{n}{N}k} \nonumber \\
&+
v_r(n,m),
\label{eq:2d_received_scalar}
\end{align}
where $\alpha_{r,t,k,i}$ denotes the $2$D scattering coefficient corresponding to the $k$th Doppler grid point and the $i$th delay grid point. Define the Doppler and delay steering matrices as $\mathbf{B}_{\nu,t}=[\boldsymbol{\beta}_{\nu,t}(:,0),\ldots,\boldsymbol{\beta}_{\nu,t}(:,D_{\nu}-1)]\in\mathbb{C}^{N\times D_{\nu}}$ and $\mathbf{B}_{\tau,t}=[\boldsymbol{\beta}_{\tau,t}(:,0),\ldots,\boldsymbol{\beta}_{\tau,t}(:,D_{\tau}-1)]\in\mathbb{C}^{M\times D_{\tau}}$, respectively, where $\boldsymbol{\beta}_{\nu,t}(n,k)=x_{\nu,t}[n]e^{j2\pi nk/N}$ and $\boldsymbol{\beta}_{\tau,t}(m,i)=x_{\tau,t}^{*}[m]e^{-j2\pi mi/M}$. Let $\mathbf{A}_{r,t}\in\mathbb{C}^{D_{\nu}\times D_{\tau}}$ denote the $2$D scattering coefficient matrix for the link between the $t$th transmit LED and the $r$th receive PD. The received signal matrix is then given by
\vspace{-2mm}
\begin{align}
\mathbf{Y}_r
=
\sum_{t=1}^{N_t}
\mathbf{B}_{\nu,t}
\mathbf{A}_{r,t}
\mathbf{B}_{\tau,t}^{T}
+
\mathbf{V}_r .
\label{eq:2d_matrix_model}
\end{align}
For compact sparse representation, define $\mathbf{D}_{r,t}=\mathbf{A}_{r,t}\mathbf{B}_{\tau,t}^{T}\in\mathbb{C}^{D_{\nu}\times M}$. Then, \eqref{eq:2d_matrix_model} can be equivalently expressed as $\mathbf{Y}_r=\sum_{t=1}^{N_t}\mathbf{B}_{\nu,t}\mathbf{D}_{r,t}+\mathbf{V}_r$. Vectorizing the \eqref{eq:2d_matrix_model}, yields
\vspace{-2mm}
\begin{align}
\bar{\mathbf{y}}_r
=
\sum_{t=1}^{N_t}
\left(
\mathbf{I}_{M}
\otimes
\mathbf{B}_{\nu,t}
\right)
\bar{\mathbf{d}}_{r,t}
+
\bar{\mathbf{z}}_r,
\label{eq:2d_vectorized_model}
\end{align}
where $\bar{\mathbf{y}}_r=\operatorname{vec}(\mathbf{Y}_r)$, $\bar{\mathbf{d}}_{r,t}=\operatorname{vec}(\mathbf{D}_{r,t})$, $\bar{\mathbf{z}}_r=\operatorname{vec}(\mathbf{V}_r)$, $\mathbf{I}_{M}$ is the $M\times M$ identity matrix, and $\otimes$ denotes the Kronecker product. By defining $\boldsymbol{\Psi}_{t}=\mathbf{I}_{M}\otimes\mathbf{B}_{\nu,t}$, the vectorized received signal becomes $\bar{\mathbf{y}}_r=\sum_{t=1}^{N_t}\boldsymbol{\Psi}_{t}\bar{\mathbf{d}}_{r,t}+\bar{\mathbf{z}}_r$.
Stacking the contributions from all transmit LEDs gives $\boldsymbol{\Psi}_{\mathrm{MIMO}}=[\boldsymbol{\Psi}_{1},\boldsymbol{\Psi}_{2},\ldots,\boldsymbol{\Psi}_{N_t}]$ and $\bar{\mathbf{d}}_r=[\bar{\mathbf{d}}_{r,1}^{T},\bar{\mathbf{d}}_{r,2}^{T},\ldots,\bar{\mathbf{d}}_{r,N_t}^{T}]^{T}$. Hence, the received signal for the $r$th PD is compactly written as $\bar{\mathbf{y}}_r=\boldsymbol{\Psi}_{\mathrm{MIMO}}\bar{\mathbf{d}}_r+\bar{\mathbf{z}}_r$. Finally, stacking the observations from all $N_r$ receive PDs yields
\vspace{-2mm}
\begin{align}
\mathbf{Y}
=
\boldsymbol{\Psi}_{\mathrm{MIMO}}
\mathbf{D}
+
\mathbf{Z},
\label{eq:2d_mimo_sparse_model}
\end{align}
where $\mathbf{Y}=[\bar{\mathbf{y}}_1,\bar{\mathbf{y}}_2,\ldots,\bar{\mathbf{y}}_{N_r}]$, $\mathbf{D}=[\bar{\mathbf{d}}_1,\bar{\mathbf{d}}_2,\ldots,\bar{\mathbf{d}}_{N_r}]$, and $\mathbf{Z}=[\bar{\mathbf{z}}_1,\bar{\mathbf{z}}_2,\ldots,\bar{\mathbf{z}}_{N_r}]$. The model in \eqref{eq:2d_mimo_sparse_model} forms the basis for the proposed on-grid and off-grid $2$D sparse sensing PE.
\vspace{-4mm}

\subsection{$2$D off-grid sparse representation for sensing PE}

In this subsection, the $2$D on-grid formulation is extended to an $2$D off-grid sensing PE model by jointly accounting for delay and Doppler mismatches. The proposed formulation follows the $1$D off-grid framework while exploiting the separable structure of the DD-domain dictionary. Let $\bar{\boldsymbol{\tau}}=[\bar{\tau}_0,\bar{\tau}_1,\ldots,\bar{\tau}_{D_\tau-1}]^T$ and $\bar{\boldsymbol{\nu}}=[\bar{\nu}_0,\bar{\nu}_1,\ldots,\bar{\nu}_{D_\nu-1}]^T$ denote the predefined uniform delay and Doppler grids, respectively. These grids are assumed to be sufficiently dense such that the true target parameters lie close to their nearest grid points. Using these grids, the received signal at the $r$th PD can be approximated as
\vspace{-2mm}
\begin{align}
y_r(n,m)
=
\sum_{t=1}^{N_t}
\sum_{i=0}^{D_\tau-1}
\sum_{k=0}^{D_\nu-1}
\alpha_{r,t,k,i}
\beta_{\nu,t,k}[n]
\beta_{\tau,t,i}[m]
+
v_r(n,m),
\label{eq:2d_offgrid_scalar}
\end{align}
where $\beta_{\nu,t,k}[n]=x_{\nu,t}[n]e^{j2\pi nk/N}$ and $\beta_{\tau,t,i}[m]=x_{\tau,t}^{*}[m]e^{-j2\pi mi/M}$ denote the Doppler and delay steering components, respectively.

%\subsubsection{Off-grid modeling via first-order approximation}
In practice, the true delay and Doppler parameters may not exactly coincide with the predefined grid points. Therefore, they are modeled as $\tilde{\tau}_i=\bar{\tau}_i+\epsilon_{\tau,i}$ and $\tilde{\nu}_k=\bar{\nu}_k+\epsilon_{\nu,k}$, where $\epsilon_{\tau,i}\in[-r_\tau/2,r_\tau/2]$ and $\epsilon_{\nu,k}\in[-r_\nu/2,r_\nu/2]$ denote the off-grid deviations. Here, $r_\tau=2\tau_{\max}/D_\tau$ and $r_\nu=2\nu_{\max}/D_\nu$ are the delay and Doppler grid resolutions, respectively. Using a first-order Taylor approximation around the nearest grid points, the delay and Doppler steering components are approximated as $\beta_{\tau,t,i}[m]\approx \beta_{\tau,t,i}(\bar{\tau}_i)[m]+\epsilon_{\tau,i}\beta'_{\tau,t,i}[m]$ and $\beta_{\nu,t,k}[n]\approx \beta_{\nu,t,k}(\bar{\nu}_k)[n]+\epsilon_{\nu,k}\beta'_{\nu,t,k}[n]$, where
\vspace{-2mm}
\begin{align}
\beta'_{\tau,t,i}[m]
&=-j2\pi m\Delta f
x_{\tau,t}^{*}[m]e^{-j2\pi m\Delta f\bar{\tau}_i},
\notag \\
\beta'_{\nu,t,k}[n]
&=j2\pi nT_s
x_{\nu,t}[n]e^{j2\pi nT_s\bar{\nu}_k}.
\label{eq:2d_offgrid_derivatives}
\end{align}
These derivative terms capture the first-order perturbations caused by delay and Doppler basis mismatch.

%\subsubsection{$2$D off-grid SSR formulation}
Let $\mathbf{B}_{\nu,t}$ and $\mathbf{B}_{\tau,t}$ denote the on-grid Doppler and delay dictionaries, respectively, and let $\mathbf{B}'_{\nu,t}$ and $\mathbf{B}'_{\tau,t}$ denote their corresponding derivative dictionaries. Incorporating the off-grid corrections, the effective dictionaries are defined as $\tilde{\mathbf{B}}_{\nu,t}=\mathbf{B}_{\nu,t}+\mathbf{B}'_{\nu,t}\operatorname{diag}(\boldsymbol{\epsilon}_{\nu})$ and $\tilde{\mathbf{B}}_{\tau,t}=\mathbf{B}_{\tau,t}+\mathbf{B}'_{\tau,t}\operatorname{diag}(\boldsymbol{\epsilon}_{\tau})$. Hence, the received signal at the $r$th receive PD is given by
\vspace{-2mm}
\begin{align}
\mathbf{Y}_r
=
\sum_{t=1}^{N_t}
\tilde{\mathbf{B}}_{\nu,t}
\mathbf{A}_{r,t}
\tilde{\mathbf{B}}_{\tau,t}^{T}
+
\mathbf{V}_r,
\label{eq:2d_offgrid_matrix}
\end{align}
where $\mathbf{A}_{r,t}\in\mathbb{C}^{D_\nu\times D_\tau}$ is the $2$D scattering coefficient matrix corresponding to the $t$th transmit LED and the $r$th receive PD.
Vectorizing \eqref{eq:2d_offgrid_matrix}, the received signal can be expressed as
\vspace{-2mm}
\begin{align}
\mathbf{y}_r
=
\sum_{t=1}^{N_t}
\left(
\tilde{\mathbf{B}}_{\tau,t}
\otimes
\tilde{\mathbf{B}}_{\nu,t}
\right)
\operatorname{vec}(\mathbf{A}_{r,t})
+
\mathbf{v}_r,
\label{eq:2d_offgrid_vectorized}
\end{align}
where $\otimes$ denotes the Kronecker product, $\mathbf{y}_r=\operatorname{vec}(\mathbf{Y}_r)$, and $\mathbf{v}_r=\operatorname{vec}(\mathbf{V}_r)$. By defining the effective MIMO dictionary $\tilde{\boldsymbol{\Psi}}_{\mathrm{MIMO}}$ and the stacked sparse coefficient vector $\tilde{\mathbf{d}}_r$, \eqref{eq:2d_offgrid_vectorized} can be compactly written as $\mathbf{y}_r
=
\tilde{\boldsymbol{\Psi}}_{\mathrm{MIMO}}
\tilde{\mathbf{d}}_r
+
\mathbf{v}_r.$
% \begin{align}
% \mathbf{y}_r
% =
% \tilde{\boldsymbol{\Psi}}_{\mathrm{MIMO}}
% \tilde{\mathbf{d}}_r
% +
% \mathbf{v}_r.
% \label{eq:2d_offgrid_compact}
% \end{align}
Finally, stacking the observations from all $N_r$ receive PDs yields the global measurement model
\vspace{-2mm}
\begin{align}
\mathbf{Y}
=
\tilde{\boldsymbol{\Psi}}_{\mathrm{MIMO}}
\tilde{\mathbf{D}}
+
\mathbf{V},
\label{eq:2d_offgrid_global}
\end{align}
where $\mathbf{Y}=[\mathbf{y}_1,\mathbf{y}_2,\ldots,\mathbf{y}_{N_r}]$, $\tilde{\mathbf{D}}=[\tilde{\mathbf{d}}_1,\tilde{\mathbf{d}}_2,\ldots,\tilde{\mathbf{d}}_{N_r}]$, and $\mathbf{V}=[\mathbf{v}_1,\mathbf{v}_2,\ldots,\mathbf{v}_{N_r}]$. The model in \eqref{eq:2d_offgrid_global} provides the basis for $2$D off-grid sparse recovery, where the on-grid support identifies the coarse DD locations, while $\boldsymbol{\epsilon}_{\tau}$ and $\boldsymbol{\epsilon}_{\nu}$ refine the corresponding delay and Doppler estimates.

\vspace{-2mm}
\subsection{$2$D off-grid BL ($2$D-OffGBL)-based target PE}

Based on the derived $2$D off-grid SSR model, a two-stage BL-based sensing PE algorithm is proposed. In the first stage, the sparse coefficient matrix $\tilde{\mathbf{D}}$ is estimated by exploiting the joint sparsity across the receive PDs. For the $r$th receive PD and the $t$th transmit LED, the corresponding coefficient vector is defined as $\tilde{\mathbf{d}}_{r,t}=\operatorname{vec}(\mathbf{A}_{r,t}\mathbf{B}_{\tau,t}^{T})$, and the full coefficient vector $\tilde{\mathbf{d}}_r$ is obtained by stacking the contributions from all transmit LEDs. In the second stage, the scattering matrices $\mathbf{A}_{r,t}$ are recovered from $\tilde{\mathbf{D}}$, and the DD estimates are refined using the estimated off-grid deviations. This two-stage strategy enables joint estimation of sparse on-grid components and off-grid corrections with reduced computational complexity.

Similar to the $1$D-OffGBL formulation, a parameterized Gaussian prior is imposed on the $g$th row $\tilde{\boldsymbol{\delta}}_g$ of $\tilde{\mathbf{D}}$ as \cite{1315936}
\vspace{-2mm}
\begin{align}
p(\tilde{\boldsymbol{\delta}}_g;\gamma_g)
=
\mathcal{CN}
\left(
\mathbf{0}_{N_r\times 1},
\gamma_g^{-1}\mathbf{I}_{N_r}
\right)
=
\prod_{r=1}^{N_r}
\frac{\gamma_g}{\pi}
\exp\left(
-\gamma_g|\tilde{d}_{g,r}|^2
\right),
\label{eq:2d_sbl_prior}
\end{align}
where $\gamma_g$ is the precision hyperparameter controlling the sparsity of the $g$th row, $0\le g\le G-1$, and $G=MD_\nu N_t$. The prior distribution of $\tilde{\mathbf{D}}$ is given by $p(\tilde{\mathbf{D}};\boldsymbol{\gamma})=\prod_{g=0}^{G-1}p(\tilde{\boldsymbol{\delta}}_g;\gamma_g)$, where $\boldsymbol{\gamma}=[\gamma_0,\gamma_1,\ldots,\gamma_{G-1}]^T$. To regularize the estimation, a Gamma prior is assigned to $\boldsymbol{\gamma}$ as \cite{1315936}
\vspace{-2mm}
\begin{align}
p(\boldsymbol{\gamma})
=
\prod_{g=0}^{G-1}
\mathrm{Gamma}(\gamma_g|c+1,d)
=
\prod_{g=0}^{G-1}
\frac{d^c\gamma_g^c e^{-d\gamma_g}}
{\Gamma(c+1)}.
\label{eq:2d_gamma_prior}
\end{align}
Let $\boldsymbol{\Omega}_d=\{\tilde{\mathbf{D}},\boldsymbol{\gamma},\tilde{\boldsymbol{\epsilon}}_\nu\}$ denote the set of hidden variables. The MAP estimation problem is formulated as \cite{7547360}
\vspace{-2mm}
\begin{align}
\widehat{\boldsymbol{\Omega}}_d
=
\arg\max_{\boldsymbol{\Omega}_d}
\ln p(\mathbf{Y},\boldsymbol{\Omega}_d).
\label{eq:2d_map_problem}
\end{align}
The likelihood of the observation is modeled as $p(\mathbf{Y}|\tilde{\mathbf{D}},\boldsymbol{\gamma},\tilde{\boldsymbol{\epsilon}}_\nu)=\mathcal{CN}(\mathbf{Y}|\tilde{\boldsymbol{\Psi}}_{\mathrm{MIMO}}\tilde{\mathbf{D}},\sigma^2\mathbf{I})$. Accordingly, the posterior distribution of the $r$th column of $\tilde{\mathbf{D}}$ follows \cite{1315936}
$q(\tilde{\boldsymbol{\delta}}_r)=\mathcal{CN}(\boldsymbol{\mu}_r^{(k+1)},\boldsymbol{\Sigma}^{(k+1)})$, where
\vspace{-2mm}
\begin{align}
\boldsymbol{\Sigma}^{(k+1)}
&=
\left(
\frac{1}{\sigma^2}
\tilde{\boldsymbol{\Psi}}_{\mathrm{MIMO}}^{H}
\tilde{\boldsymbol{\Psi}}_{\mathrm{MIMO}}
+
\boldsymbol{\Gamma}^{(k+1)}
\right)^{-1}, \notag \\
\mathbf{U}^{(k+1)}
&=
\frac{1}{\sigma^2}
\boldsymbol{\Sigma}^{(k+1)}
\tilde{\boldsymbol{\Psi}}_{\mathrm{MIMO}}^{H}
\mathbf{Y},
\label{eq:2d_posterior_update}
\end{align}
with $\boldsymbol{\Gamma}=\operatorname{diag}(\boldsymbol{\gamma})$ and $\mathbf{U}^{(k+1)}=[\boldsymbol{\mu}_1^{(k+1)},\ldots,\boldsymbol{\mu}_{N_r}^{(k+1)}]$. These updates follow the same Bayesian posterior derivation used for the $1$D-OffGBL estimator. The hyperparameter associated with the $g$th row is updated as \cite{7547360}
\vspace{-2mm}
\begin{align}
\gamma_g^{(k+1)}
=
\frac{c+N_r}
{d+\sum_{r=1}^{N_r}
\left(
|\mu_{g,r}|^2+\Sigma_{g,g}
\right)},
\quad
0\le g\le G-1.
\label{eq:2d_gamma_update}
\end{align}
The Doppler off-grid deviation vector is updated as
\vspace{-2mm}
\begin{align}
\tilde{\boldsymbol{\epsilon}}_\nu^{(k+1)}
=
\left(
\Re\left\{
\boldsymbol{\Psi}_{\mathrm{MIMO},\nu}^{H}
\boldsymbol{\Psi}_{\mathrm{MIMO},\nu}
\odot
\left(
\mathbf{U}\mathbf{U}^{H}
+
N_r\boldsymbol{\Sigma}
\right)
\right\}
\right)^{\dagger}
\mathbf{p}_\nu.
\label{eq:2d_doppler_offgrid_update}
\end{align}
%where $(\cdot)^{\dagger}$ denotes the Moore-Penrose pseudo-inverse, $\odot$ is the Hadamard product, and $\Re\{\cdot\}$ extracts the real part. 
The auxiliary vector $\mathbf{p}_\nu$ is given by
\vspace{-2mm}
\begin{align}
\mathbf{p}_{\nu}
&=
\Re\left\{
\sum_{r=1}^{N_r}
\operatorname{diag}(\boldsymbol{\mu}_{r})
\boldsymbol{\Psi}_{\mathrm{MIMO},\nu}^{H}
\left(
\mathbf{y}_{r}
-
\boldsymbol{\Psi}_{\mathrm{MIMO},\nu}
\boldsymbol{\mu}_{r}
\right)
\right\}
\notag \\
&\quad
-
N_r
\Re\left\{
\operatorname{diag}
\left(
\boldsymbol{\Psi}_{\mathrm{MIMO},\nu}^{H}
\boldsymbol{\Psi}_{\mathrm{MIMO},\nu}
\boldsymbol{\Sigma}
\right)
\right\},
\label{eq:2d_p_nu}
\end{align}
where $\boldsymbol{\Psi}_{\mathrm{MIMO},\nu}=\tilde{\boldsymbol{\Psi}}_{\mathrm{MIMO}}'(\tilde{\boldsymbol{\nu}})$ denotes the derivative of the effective MIMO dictionary with respect to the Doppler grid. The Doppler grid is then refined as
\vspace{-2mm}
\begin{align}
\tilde{\boldsymbol{\nu}}^{(k+1)}
=
\tilde{\boldsymbol{\nu}}^{(k)}
+
\tilde{\boldsymbol{\epsilon}}_\nu^{(k+1)}.
\label{eq:2d_doppler_grid_update}
\end{align}
The above steps are repeated until convergence. Since the hyperparameters determine the row sparsity of $\widehat{\tilde{\mathbf{D}}}$, the final estimate is further refined by pruning rows associated with large precision values \cite{8718676}. For a chosen threshold $\gamma_{\mathrm{th}}$, the pruned estimate is obtained as
\vspace{-1mm}
\begin{align}
\widehat{\tilde{\mathbf{D}}}(g,:)
=
\begin{cases}
\widehat{\tilde{\mathbf{D}}}(g,:), & \gamma_g \le \gamma_{\mathrm{th}},\\
\mathbf{0}, & \gamma_g > \gamma_{\mathrm{th}},
\end{cases}
\label{eq:2d_pruned_D}
\end{align}
where $0\le g\le G-1$. The derivations of \eqref{eq:2d_posterior_update} to \eqref{eq:2d_p_nu} follow the same steps as those used for the $1$D-OffGBL update rules.

\begin{algorithm}[t]
\caption{$2$D-OffGBL-assisted target PE}
\label{alg:2d_sbl_offgrid}
\small
\DontPrintSemicolon

\KwIn{Received observation matrix $\mathbf{Y}\in\mathbb{C}^{MN\times N_r}$, measurement matrix $\tilde{\boldsymbol{\Psi}}_{\mathrm{MIMO}}\in\mathbb{C}^{MN\times \mathcal{D}_\nu M N_t}$, noise precision matrix $\sigma^{-2}\mathbf{I}_{\mathcal{D}_\nu M N_t}$, and stopping parameters $\epsilon$ and $t_{\max}$.}

\KwOut{Estimated $\widehat{\mathbf{A}}$, $\tilde{\boldsymbol{\tau}}$, $\tilde{\boldsymbol{\nu}}$.}

\textbf{Initialization:} Set $\widehat{\gamma}_j^{(0)}=1$ for all $1\le j\le \mathcal{D}_\nu M N_t$, $\widehat{\boldsymbol{\Gamma}}^{(0)}=\mathbf{I}_{\mathcal{D}_\nu M N_t}$.
Set $\boldsymbol{\gamma}^{(-1)}=\mathbf{0}$, $\widehat{\boldsymbol{\Gamma}}^{(-1)}=\mathbf{0}$, and $t=0$.
Initialize $\boldsymbol{\epsilon}_{\tau}^{(0)}=\mathbf{0}$ and $\boldsymbol{\epsilon}_{\nu}^{(0)}$ %as given in Section~VI-A
%\tcp{Apply row-wise separately, if required}

\While{$\left\lVert \boldsymbol{\gamma}^{(t)}-\boldsymbol{\gamma}^{(t-1)}\right\rVert_2 > \epsilon$ \text{and} $t<t_{\max}$}{
    
    \textbf{E-step:} Compute the \textit{a posteriori} covariance and mean
    \vspace{-3mm}
    \[
    \begin{aligned}
    \boldsymbol{\Sigma}^{(t+1)}
    &=
    \left(
    \frac{1}{\sigma^{2}}
    \tilde{\boldsymbol{\Psi}}_{\mathrm{MIMO}}^{H}
    \tilde{\boldsymbol{\Psi}}_{\mathrm{MIMO}}
    +
    \boldsymbol{\Gamma}^{(t)}
    \right)^{-1},\\
    \mathbf{U}^{(t+1)}
    &=
    \left[
    \boldsymbol{\mu}_{1}^{(t+1)},
    \ldots,
    \boldsymbol{\mu}_{N_r}^{(t+1)}
    \right]
    =
    \frac{1}{\sigma^{2}}
    \boldsymbol{\Sigma}^{(t+1)}
    \tilde{\boldsymbol{\Psi}}_{\mathrm{MIMO}}^{H}
    \mathbf{Y}.
    \end{aligned}
    \]

    \textbf{M-step:}

    \textbf{(i) Update hyperparameter vector $\boldsymbol{\gamma}^{(t+1)}$:}\\
    \For{$g=1,2,\ldots,\mathcal{D}_\nu M N_t$}{\vspace{-1mm}
        \[
        \begin{aligned}
        \gamma_g^{(t+1)}
        &=
        \frac{c+N_r}
        {\,d+\sum_{r=1}^{N_r}
        \Upsilon_r\left(
        \boldsymbol{\epsilon}_{\nu}^{(t)}
        \right)},\\   
    \boldsymbol{\gamma}^{(t+1)}
    &=
    \left[
    \gamma_1^{(t+1)},\ldots,\gamma_{\mathcal{D}_\nu M N_t}^{(t+1)}
    \right]^T,\\
    \boldsymbol{\Gamma}^{(t+1)}
    &=
    \operatorname{diag}\left(\boldsymbol{\gamma}^{(t+1)}\right).
    \end{aligned}
    \]
    }
    \textbf{(ii) Update off-grid parameters:}
    Update $\mathbf{p}_{\nu}$ according to \eqref{eq:2d_p_nu}
    Update $\tilde{\boldsymbol{\epsilon}}_{\nu}^{(t+1)}$ according to \eqref{eq:2d_doppler_offgrid_update}

    Set $t\leftarrow t+1$
}

\textbf{return:} $\widehat{\tilde{\mathbf{D}}}=\mathbf{U}^{(t)}.$
Recover $\mathbf{d}_{r,t}$ from each column of $\widehat{\tilde{\mathbf{D}}}$,
reshape each recovered vector to obtain $\mathbf{D}_{r,t}$,
estimate $\mathbf{A}_{r,t}$, and $\widehat{\tilde{\boldsymbol{\tau}}}^{(t)}$ using Algorithm~1.
Set $\widehat{\mathbf{A}}=\left[\mathbf{A}_{r,t}\right]$.
\end{algorithm}
\vspace{-2mm}
\section{CE, range, and velocity estimation}
This section presents the $1$D and $2$D sparse channel representation and estimation, followed by the range and velocity estimation for both $1$D and $2$D models using the target PE updates discussed in the previous sections for the considered framework. 
\vspace{-3mm}
\subsection{$1$D sparse representation and CE}
To formulate the downlink MIMO DCO-OFDM VLC CE problem in the DD-domain, let $L_\tau$ and $L_\nu$ denote the delay and Doppler grid sizes, respectively. The corresponding grids are defined as $\mathcal{L}_\tau=\{\tau_i=i/(M\Delta f)\}_{i=0}^{L_\tau-1}$ and $\mathcal{L}_\nu=\{\nu_k=k/(NT_s)\}_{k=0}^{L_\nu-1}$. Using these grids, the received downlink signal at the $r$th UE PD can be expressed as
\vspace{-2mm}
\begin{align}
y_r(n,m)
=
\sum_{t=1}^{N_t}
\sum_{i=0}^{L_\tau-1}
\sum_{k=0}^{L_\nu-1}
\xi_{r,t,i,k}
\beta_{i,k,t}(n,m)
+
z_r(n,m),
\label{eq:1d_sparse_ce_scalar}
\end{align}
where $\xi_{r,t,i,k}$ denotes the channel coefficient associated with the $i$th delay grid point and the $k$th Doppler grid point for the link between the $t$th transmit LED and the $r$th UE PD. Further, $\beta_{i,k,t}(n,m)=x_t(n,m)e^{-j2\pi mi/M}e^{j2\pi nk/N}$ and $z_r(n,m)$ denotes additive noise. Following the same vectorization procedure used for the $1$D sparse radar model, the CE problem can be written as
\vspace{-3mm}
\begin{align}
\mathbf{Y}_{\mathrm{c}}
=
\boldsymbol{\Psi}_{\mathrm{MIMO},\mathrm{c}}
\mathbf{H}
+
\mathbf{Z},
\label{eq:1d_sparse_ce_matrix}
\end{align}
where $\mathbf{Y}_{\mathrm{c}}$ denotes the stacked received communication signal, $\boldsymbol{\Psi}_{\mathrm{MIMO},\mathrm{c}}$ is the communication sensing dictionary, $\mathbf{H}$ collects the sparse channel coefficients, and $\mathbf{Z}$ is the noise matrix. The coefficient vectors are sparse since the number of dominant channel scatterers $L$ satisfies $L\ll L_\tau L_\nu$. Therefore, the $1$D on-grid and off-grid sparse recovery frameworks and $1$D-OffGBL algorithm developed for sensing PE can be directly adapted for sparse CE.
\vspace{-4mm}
\subsection{$2$D sparse representation and CE}
For the $2$D sparse CE formulation, the pilot matrix is assumed to be separable across the Doppler and delay dimensions, i.e., $\mathbf{X}=\mathbf{x}_{\nu}\mathbf{x}_{\tau}^{H}$, where $\mathbf{X}\in\mathbb{C}^{N\times M}$, $\mathbf{x}_{\nu}\in\mathbb{C}^{N\times 1}$, and $\mathbf{x}_{\tau}\in\mathbb{C}^{M\times 1}$ collect the pilot symbols across the time and frequency domains, respectively \cite{11039779}. Under this separable structure, the received signal can be written as
\vspace{-2mm}
\begin{align}
y_r(n,m)
&=
\sum_{t=1}^{N_t}
\sum_{i=0}^{L_\tau-1}
\sum_{k=0}^{L_\nu-1}
\xi_{r,t,k,i}
x_{\nu,t}[n]x_{\tau,t}^{*}[m]
e^{-j2\pi mi/M} \nonumber \\
& \times e^{j2\pi nk/N}
+
z_r(n,m).
\label{eq:2d_sparse_ce_scalar}
\end{align}
Since \eqref{eq:2d_sparse_ce_scalar} has the same structure as the $2$D sparse sensing observation model, the corresponding vectorized CE model can be obtained by following the same matrix construction and stacking procedure. Thus, the $2$D sparse CE model is 
\vspace{-2mm}
\begin{align}
\mathbf{Y}_{\mathrm{c}}
=
\boldsymbol{\Psi}_{\mathrm{MIMO},\mathrm{c}}
\mathbf{H}
+
\mathbf{Z},
\label{eq:2d_sparse_ce_matrix}
\end{align}
where $\boldsymbol{\Psi}_{\mathrm{MIMO},\mathrm{c}}$ is the $2$D communication dictionary and $\mathbf{H}$ contains the sparse channel coefficients. The model in \eqref{eq:2d_sparse_ce_matrix} serves as the basis for both $2$D on-grid and off-grid sparse recovery framework and the $2$D-OffGBL-based CE algorithm. %Moreover, the estimated DD support can also be used to infer range and velocity information through the estimated delay and Doppler components.

\vspace{-4mm}
\subsection{Range and velocity estimation}
\vspace{-1mm}
\subsubsection{$1$D range and velocity estimation}
The scattering scene can be reconstructed from the estimated row-sparse RCS coefficient matrix $\widehat{\mathbf{A}}$ obtained using the BL-based estimator. For the $1$D sparse model, the received signal is expressed as $\mathbf{Y}=\boldsymbol{\Phi}_{\mathrm{MIMO}}\mathbf{A}+\mathbf{Z}$, where $\boldsymbol{\Phi}_{\mathrm{MIMO}}\in\mathbb{C}^{MN\times D_\tau D_\nu N_t}$ and $\mathbf{A}\in\mathbb{C}^{D_\tau D_\nu N_t\times N_r}$. Since $\widehat{\mathbf{A}}$ is row-sparse, only a few rows corresponding to the active DD bins contain significant entries. To detect the active targets, the averaged magnitude of the estimated RCS coefficients at the DD bin $(i,k)$ is given as \cite{11039779}
\vspace{-2mm}
\begin{align}
\bar{\alpha}(i,k)
=
\frac{1}{N_tN_r}
\sum_{t=1}^{N_t}
\sum_{r=1}^{N_r}
\left|
\widehat{\alpha}_{r,t}(i,k)
\right|.
\label{eq:avg_rcs_1d}
\end{align}
The set of detected DD indices is then defined as $\mathcal{I}=\{(i,k)\mid \bar{\alpha}(i,k)>\eta\}$, where $\eta$ is a suitably chosen detection threshold. For a detected target corresponding to $(\widehat{i},\widehat{k})\in\mathcal{I}$, the on-grid range and velocity estimates are given by
\vspace{-2mm}
\begin{align}
\widehat{R}_{p,\mathrm{on},1D}
=
\frac{c}{2M\Delta f}\widehat{i},
\qquad
\widehat{V}_{p,\mathrm{on},1D}
=
\frac{c}{2f_cNT_s}\widehat{k},
\label{eq:on_grid_range_velocity_1d}
\end{align}
where $c$ is the speed of light and $f_c$ is the carrier frequency. For off-grid estimation, the refined range and velocity estimates are obtained using the estimated off-grid deviation vectors $\widehat{\boldsymbol{\epsilon}}_{\tau}$ and $\widehat{\boldsymbol{\epsilon}}_{\nu}$. The algorithm is considered to have converged when $||\mathbf{z}^{(k+1)}-\mathbf{z}^{(k)}||_2^2<\xi$, where $\xi\ll 1$ is a predefined tolerance. After convergence, the off-grid correction terms for the detected DD bin $(\widehat{i},\widehat{k})$ are $\widehat{R}_{p,\mathrm{err},1D}=\frac{c}{2M\Delta f}\widehat{\epsilon}_{\tau}(\widehat{i})$ and $\widehat{V}_{p,\mathrm{err},1D}=\frac{c}{2f_cNT_s}\widehat{\epsilon}_{\nu}(\widehat{k})$. Thus, the refined off-grid estimates are
\vspace{-2mm}
\begin{align}
\widehat{R}_{p,\mathrm{off},1D}
&=
\frac{c}{2M\Delta f}
\left(
\widehat{i}
+
\widehat{\epsilon}_{\tau}(\widehat{i})
\right), \\
\widehat{V}_{p,\mathrm{off},1D}
&=
\frac{c}{2f_cNT_s}
\left(
\widehat{k}
+
\widehat{\epsilon}_{\nu}(\widehat{k})
\right).
\label{eq:off_grid_range_velocity_1d}
\end{align}
\vspace{-2mm}
\subsubsection{$2$D range and velocity estimation}
For the $2$D sparse formulation, the estimated coefficient matrix $\widehat{\mathbf{A}}_{r,t}$ corresponding to each transmit-receive pair contains the DD-domain scattering coefficients. A target located at the DD bin $(i,k)$ is detected by averaging the coefficient magnitudes over all transmit LEDs and receive PDs. Specifically, the detected support set is defined as \cite{11039779}
\vspace{-2mm}
\begin{align}
\mathcal{I}
=
\left\{
(i,k)
\,\middle|\,
\frac{1}{N_tN_r}
\sum_{t=1}^{N_t}
\sum_{r=1}^{N_r}
\left|
\widehat{A}_{r,t}(i,k)
\right|
>
\eta
\right\}.
\label{eq:detected_support_2d}
\end{align}
For a detected DD bin $(\widehat{i},\widehat{k})\in\mathcal{I}$, the corresponding on-grid range and velocity estimates are
\vspace{-2mm}
\begin{align}
\widehat{R}_{p,\mathrm{on},2D}
=
\frac{c}{2M\Delta f}\widehat{i},
\qquad
\widehat{V}_{p,\mathrm{on},2D}
=
\frac{c}{2f_cNT_s}\widehat{k}.
\label{eq:on_grid_range_velocity_2d}
\end{align}
After convergence of the $2$D off-grid BL algorithm, the estimated Doppler deviation vector $\widehat{\tilde{\boldsymbol{\epsilon}}}_{\nu}$ is used for velocity refinement. Similarly, after recovering the refined coefficient matrix, the delay deviation vector $\widehat{\tilde{\boldsymbol{\epsilon}}}_{\tau}$ is used for range refinement. For the detected indices $\widehat{i}$ and $\widehat{k}$, the corresponding correction terms are $\widehat{R}_{p,\mathrm{err},2D}=\frac{c}{2M\Delta f}\widehat{\tilde{\epsilon}}_{\tau}(\widehat{i})$ and $\widehat{V}_{p,\mathrm{err},2D}=\frac{c}{2f_cNT_s}\widehat{\tilde{\epsilon}}_{\nu}(\widehat{k})$. Hence, the final refined range and velocity estimates are
\vspace{-2mm}
\begin{align}
\widehat{R}_{p,\mathrm{off},2D}
&=
\frac{c}{2M\Delta f}
\left(
\widehat{i}
+
\widehat{\tilde{\epsilon}}_{\tau}(\widehat{i})
\right), \\
\widehat{V}_{p,\mathrm{off},2D}
&=
\frac{c}{2f_cNT_s}
\left(
\widehat{k}
+
\widehat{\tilde{\epsilon}}_{\nu}(\widehat{k})
\right).
\label{eq:off_grid_range_velocity_2d}
\end{align}
The vectors $\widehat{\tilde{\boldsymbol{\epsilon}}}_{\tau}$ and $\widehat{\tilde{\boldsymbol{\epsilon}}}_{\nu}$ are reshaped, if required, to extract the delay and Doppler refinements corresponding to the detected target indices.

\vspace{-5mm}
\section{Bayesian Cram\'er-Rao Bound}
This section provides the BCRBs for PE, CE, range, and velocity estimation, which show the theoretical lower bound for the above-discussed signal models.
\vspace{-5mm}
\subsection{BCRB for sensing PE}
This subsection derives the BCRB for estimating the sparse RCS coefficients for the MIMO DCO-OFDM VL-based radar framework. The received signal model is given by $\mathbf{Y}=\boldsymbol{\Phi}_{\mathrm{MIMO}}\mathbf{A}+\mathbf{V}$, where $\mathbf{V}\sim\mathcal{CN}(\mathbf{0},\sigma^2\mathbf{I})$ denotes AWGN. Let $\mathbf{A}=[\boldsymbol{\alpha}_1,\boldsymbol{\alpha}_2,\ldots,\boldsymbol{\alpha}_{N_r}]$, where $\boldsymbol{\alpha}_r\in\mathbb{C}^{G\times 1}$ where $G=D_\tau D_\nu N_t$ is the RCS coefficient vector corresponding to the $r$th receive PD. Since the noise is independent across receive PDs, the column-wise received signal is expressed as $\mathbf{y}_r=\boldsymbol{\Phi}_{\mathrm{MIMO}}\boldsymbol{\alpha}_r+\mathbf{v}_r$, for $r=1,\ldots,N_r$, where $\mathbf{v}_r\sim\mathcal{CN}(\mathbf{0},\sigma^2\mathbf{I})$. Assuming a complex Gaussian prior for $\boldsymbol{\alpha}_r$, the Bayesian Fisher information matrix (BFIM) is 
\vspace{-3mm}
\begin{align}
\mathbf{J}_{r}
=
\mathbf{J}_{0,r}
+
\mathbf{J}_{\alpha,r},
\label{eq:bfim_radar}
\end{align}
where $\mathbf{J}_{0,r}$ and $\mathbf{J}_{\alpha,r}$ denote the measurement and prior information matrices, respectively. They are defined as \cite{dosso2008bayesian}
\vspace{-2mm}
\begin{align}
\mathbf{J}_{0,r}
&=
-
\mathbb{E}_{\mathbf{y}_{r},\boldsymbol{\alpha}_{r}}
\left\{
\frac{\partial^{2}
\mathcal{L}(\mathbf{y}_{r}|\boldsymbol{\alpha}_{r})}
{\partial \boldsymbol{\alpha}_{r}\partial \boldsymbol{\alpha}_{r}^{H}}
\right\},\\
\mathbf{J}_{\alpha,r}
&=
-
\mathbb{E}_{\boldsymbol{\alpha}_{r}}
\left\{
\frac{\partial^{2}
\mathcal{L}(\boldsymbol{\alpha}_{r};\boldsymbol{\Lambda})}
{\partial \boldsymbol{\alpha}_{r}\partial \boldsymbol{\alpha}_{r}^{H}}
\right\},
\label{eq:bfim_components}
\end{align}
where $\mathcal{L}(\mathbf{y}_{r}|\boldsymbol{\alpha}_{r})=\log p(\mathbf{y}_{r}|\boldsymbol{\alpha}_{r})$ and $\mathcal{L}(\boldsymbol{\alpha}_{r};\boldsymbol{\Lambda})=\log p(\boldsymbol{\alpha}_{r};\boldsymbol{\Lambda})$. For the Gaussian likelihood and prior, these log-density functions are given by $\mathcal{L}(\mathbf{y}_{r}| \boldsymbol{\alpha}_{r})=\tilde{c}-\frac{1}{2\sigma^2}||\mathbf{y}_{r}-\boldsymbol{\Phi}_{\mathrm{MIMO}}\boldsymbol{\alpha}_{r}||_2^2$ and $\mathcal{L}(\boldsymbol{\alpha}_{r};\boldsymbol{\Lambda})=d-\frac{1}{2}\boldsymbol{\alpha}_{r}^{H}\boldsymbol{\Lambda}^{-1}\boldsymbol{\alpha}_{r}$, where $\tilde{c}$ and $d$ are constants independent of $\boldsymbol{\alpha}_r$. Hence, the BFIM for the $r$th column becomes
\vspace{-2mm}
\begin{align}
\mathbf{J}_{r}
=
\frac{1}{\sigma^{2}}
\boldsymbol{\Phi}_{\mathrm{MIMO}}^{H}
\boldsymbol{\Phi}_{\mathrm{MIMO}}
+
\boldsymbol{\Lambda}^{-1}.
\label{eq:bfim_radar_final}
\end{align}
Accordingly, the mean-squared error (MSE) lower bound for estimating $\boldsymbol{\alpha}_r$ is $\mathrm{MSE}(\widehat{\boldsymbol{\alpha}}_r)\ge \operatorname{Tr}(\mathbf{J}_r^{-1})$. Since the estimation errors across receive PDs are independent, the matrix-level BCRB for estimating $\mathbf{A}$ is
\vspace{-4mm}
\begin{align}
\mathrm{MSE}(\widehat{\mathbf{A}})
\ge
\sum_{r=1}^{N_r}
\operatorname{Tr}(\mathbf{J}_{r}^{-1}).
\label{eq:bcrb_radar_matrix}
\end{align}
\vspace{-8mm}
\subsection{BCRB for CE}
The above BCRB formulation can be directly extended to the estimation of sparse channel coefficients in the proposed BL-aided MIMO DCO-OFDM VLC framework. The received downlink signal at the UE is modeled as $\mathbf{Y}_{\mathrm{c}}=\boldsymbol{\Psi}_{\mathrm{MIMO},\mathrm{c}}\mathbf{H}+\mathbf{Z}$, where $\mathbf{H}=[\boldsymbol{\xi}_1,\boldsymbol{\xi}_2,\ldots,\boldsymbol{\xi}_{R}]$ denotes the sparse channel coefficient matrix and $\mathbf{Z}$ is AWGN. Following the same procedure as in \eqref{eq:bfim_radar} to \eqref{eq:bcrb_radar_matrix}, the MSE lower bound for estimating $\mathbf{H}$ is obtained as
\vspace{-2mm}
\begin{align}
\mathrm{MSE}(\widehat{\mathbf{H}})
\ge
\sum_{r=1}^{R}
\operatorname{Tr}\left(\mathbf{J}_{\mathrm{c},r}^{-1}\right),
\label{eq:bcrb_channel_matrix}
\end{align}
where $\mathbf{J}_{\mathrm{c},r}$ is the BFIM associated with the $r$th channel coefficient vector $\boldsymbol{\xi}_r$.
\vspace{-2mm}
\subsection{BCRB for range and velocity estimation}
This subsection derives the BCRB for estimating target range and velocity parameters using the proposed MIMO DCO-OFDM VL-based radar model. For the $r$th receive PD, the observation is written as $\mathbf{y}_r=\boldsymbol{\Phi}_{\mathrm{MIMO}}\boldsymbol{\alpha}_r+\mathbf{v}_r$. The steering atom corresponding to the $t$th transmit LED and the $s$th target is given by $\boldsymbol{\phi}_{t}(\tau_s,\nu_s)=x_t(n,m)e^{-j2\pi m\Delta f\tau_s}e^{j2\pi nT_s\nu_s}$, where $\tau_s$ and $\nu_s$ denote the delay and Doppler frequency, respectively. The mean of $\mathbf{y}_r$ is therefore
\vspace{-2mm}
\begin{align}
\boldsymbol{\mu}_r
=
\mathbb{E}\{\mathbf{y}_r\}
=
\sum_{t=1}^{N_t}
\sum_{s=1}^{S}
\alpha_{s,t,r}
\boldsymbol{\phi}_{t}(\tau_s,\nu_s).
\label{eq:mean_range_velocity}
\end{align}
Define the unknown parameter vector as $\boldsymbol{\tilde{\xi}}=[\boldsymbol{\tau}^{T},\boldsymbol{\nu}^{T}]^{T}\in\mathbb{R}^{2S\times 1}$, where $\boldsymbol{\tau}=[\tau_1,\ldots,\tau_S]^T$ and $\boldsymbol{\nu}=[\nu_1,\ldots,\nu_S]^T$. The BFIM associated with $\boldsymbol{\tilde{\xi}}$ is partitioned as \cite{kay1993statistical}
\vspace{-2mm}
\begin{align}
\mathbf{J}_{\boldsymbol{\tilde{\xi}}}
=
\begin{bmatrix}
\mathbf{J}_{\tau\tau} & \mathbf{J}_{\tau\nu}\\
\mathbf{J}_{\nu\tau} & \mathbf{J}_{\nu\nu}
\end{bmatrix},
\label{eq:fim_partition}
\end{align}
where each block is given by
\vspace{-2mm}
\begin{align}
[\mathbf{J}_{ab}]_{p,q}
=
\frac{2}{\sigma^2}
\sum_{r=1}^{N_r}
\Re\left\{
\left(
\frac{\partial \boldsymbol{\mu}_r}{\partial a_p}
\right)^H
\left(
\frac{\partial \boldsymbol{\mu}_r}{\partial b_q}
\right)
\right\},
\quad
a,b\in\{\tau,\nu\}.
\label{eq:fim_block}
\end{align}
The required derivatives are $\frac{\partial \boldsymbol{\mu}_r}{\partial \tau_s}=\sum_{t=1}^{N_t}\alpha_{s,t,r}\frac{\partial \boldsymbol{\phi}_t(\tau_s,\nu_s)}{\partial \tau_s}$ and $\frac{\partial \boldsymbol{\mu}_r}{\partial \nu_s}=\sum_{t=1}^{N_t}\alpha_{s,t,r}\frac{\partial \boldsymbol{\phi}_t(\tau_s,\nu_s)}{\partial \nu_s}$, with $\mathbf{J}_{\nu\tau}=\mathbf{J}_{\tau\nu}^{H}$.
The BCRBs for delay and Doppler estimation are obtained from 
\vspace{-2mm}
$\mathbf{J}_{\boldsymbol{\tilde{\xi}}}^{-1}$ as 
\begin{align}
\mathrm{BCRB}(\boldsymbol{\tau})
=
[\mathbf{J}_{\boldsymbol{\tilde{\xi}}}^{-1}]_{1:S,1:S},
~~
\mathrm{BCRB}(\boldsymbol{\nu})
=
[\mathbf{J}_{\boldsymbol{\tilde{\xi}}}^{-1}]_{S+1:2S,S+1:2S}.
\label{eq:bcrb_tau_nu}
\end{align}
Thus, the corresponding MSE lower bounds are $\mathrm{MSE}(\widehat{\boldsymbol{\tau}})\ge \operatorname{Tr}([\mathbf{J}_{\boldsymbol{\tilde{\xi}}}^{-1}]_{1:S,1:S})$ and $\mathrm{MSE}(\widehat{\boldsymbol{\nu}})\ge \operatorname{Tr}([\mathbf{J}_{\boldsymbol{\tilde{\xi}}}^{-1}]_{S+1:2S,S+1:2S})$.
The physical range and velocity vectors are related to delay and Doppler as $\mathbf{R}=\frac{c}{2}\boldsymbol{\tau}$ and $\mathbf{V}=\frac{c}{2f_c}\boldsymbol{\nu}$. Defining the Jacobian matrix as
\vspace{-3mm}
\begin{align}
\mathbf{T}
=
\frac{\partial(\mathbf{R},\mathbf{V})}
{\partial(\boldsymbol{\tau},\boldsymbol{\nu})}
=
\begin{bmatrix}
\frac{c}{2}\mathbf{I}_S & \mathbf{0}\\
\mathbf{0} & \frac{c}{2f_c}\mathbf{I}_S
\end{bmatrix},
\label{eq:jacobian_range_velocity}
\end{align}
the BCRB for $(\mathbf{R},\mathbf{V})$ can be obtained through the covariance transformation as
\vspace{-3mm}
\begin{align}
\mathrm{BCRB}(\mathbf{R},\mathbf{V})
=
\mathbf{T}
\mathbf{J}_{\boldsymbol{\tilde{\xi}}}^{-1}
\mathbf{T}^{T}.
\label{eq:bcrb_range_velocity_full}
\end{align}
Consequently, the range and velocity BCRBs are $\mathrm{BCRB}(\mathbf{R})
=
[\mathrm{BCRB}(\mathbf{R},\mathbf{V})]_{1:S,1:S}$ and $\mathrm{BCRB}(\mathbf{V})
=
[\mathrm{BCRB}(\mathbf{R},\mathbf{V})]_{S+1:2S,S+1:2S}.$
% \begin{align}
% \mathrm{BCRB}(\mathbf{R})
% =
% [\mathrm{BCRB}(\mathbf{R},\mathbf{V})]_{1:S,1:S}, \notag \\
% \mathrm{BCRB}(\mathbf{V})
% =
% [\mathrm{BCRB}(\mathbf{R},\mathbf{V})]_{S+1:2S,S+1:2S}.
% \label{eq:bcrb_range_velocity}
% \end{align}

% \begin{table*}[t]
% \centering
% \caption{Simulation Parameters}
% \label{tab:sim_params}
% \resizebox{!}{!}{%
% \begin{tabular}{|l|c|l|c|}
% \hline
% \textbf{Parameter} & \textbf{Value} & \textbf{Parameter} & \textbf{Value} \\ 
% \hline

% % System Bandwidth ($B$) & $$ 
% % & Subcarrier Spacing ($\Delta{f}$) & $$ \\

% No.of transmitter LEDs at BS ($N_t$) & $4$ 
% & No.of sensing PDs at BS & $4$ \\

% No.of receiver PDs at the communication user & $2$ 
% & Modulation Scheme & DCO-OFDM \\

% DC Bias of DCO-OFDM  & $7~\mathrm{dB}$ 
% & Constellation Scheme & $16$-QAM \\

% Number of subcarriers ($M$) & $16$ 
% & Length of the cyclic prefix ($N_{\mathrm{CP}}$) & $4$ \\

% No. of scatterers ($L$) & $5$ 
% & No. of targets, clutters ($P,N_\mathrm{cl}$) & $4,1$ \\

% Delay bins for sensing ($D_{\tau}$) & $16$ 
% & Doppler bins for sensing ($D_{\nu}$) & $8$ \\

% No. of DCO-OFDM symbols ($N$) & $8$ 
% & Monte-Carlo simulations & $100$ \\

% \hline
% \end{tabular}%
% }
% \end{table*}
\vspace{-4mm}
\section{Simulation Results}

This section evaluates the performance of the proposed $1$D-OffGBL and $2$D-OffGBL techniques for target PE and CE in the considered MIMO DCO-OFDM VL-ISAC system. Both $1$D and $2$D VL-ISAC MIMO configurations are examined, and the simulation parameters are summarized in Table \ref{tab:sim_params}. The targets are randomly distributed in the scattering environment, and their RCS coefficients are generated as $\alpha_s\sim\mathcal{CN}(0,1)$. %Similarly, the communication channel gains are modeled as $\xi_l\sim\mathcal{N}(0,1)$. 
For both the $1$D and $2$D BL-based estimators, the stopping tolerance is set to $10^{-6}$, and the maximum number of iterations is fixed at $100$.
\begin{figure*}[t]
\centering
\captionsetup[subfigure]{justification=centering}
\subfloat[]{\label{nmse1}\includegraphics[width=60mm,height=50mm]{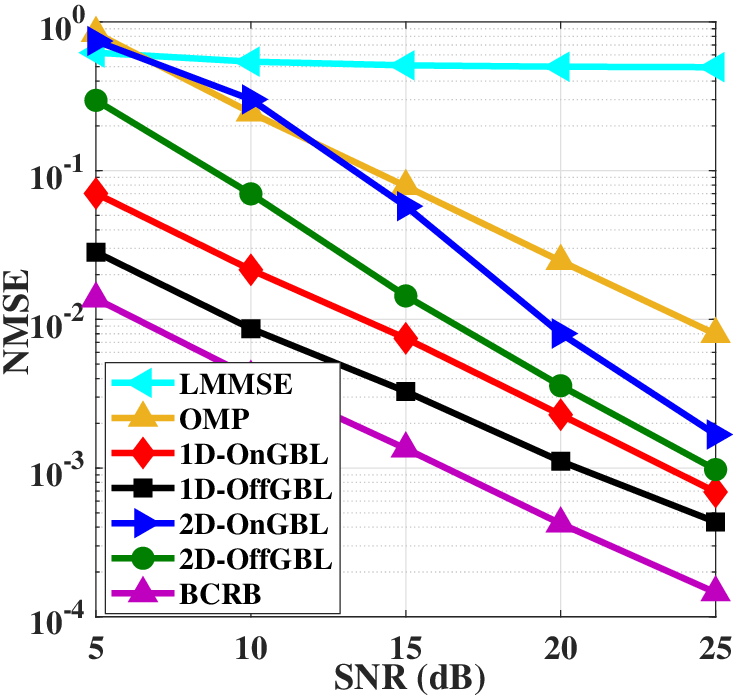}}
\subfloat[]{\label{nmse2}\includegraphics[width=60mm,height=50mm]{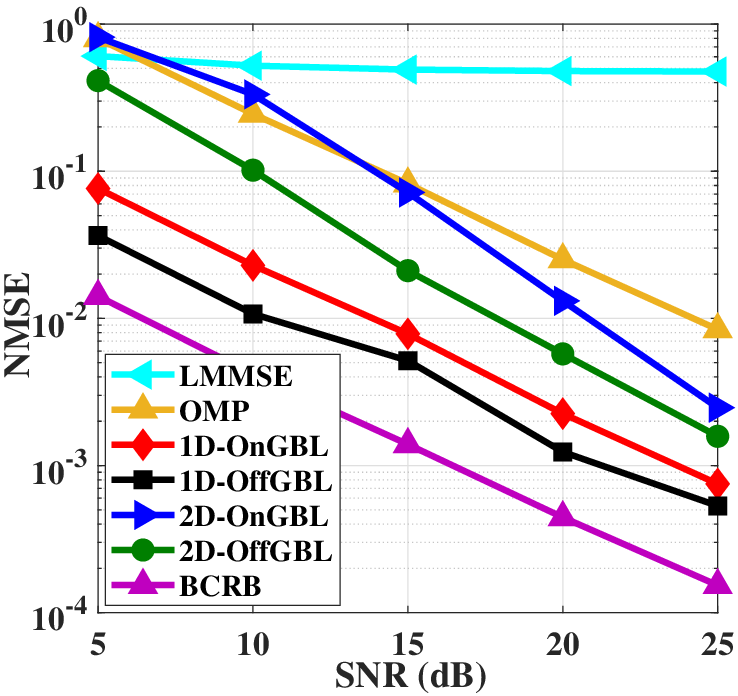}}
\subfloat[]{\label{nmse3}\includegraphics[width=60mm,height=50mm]{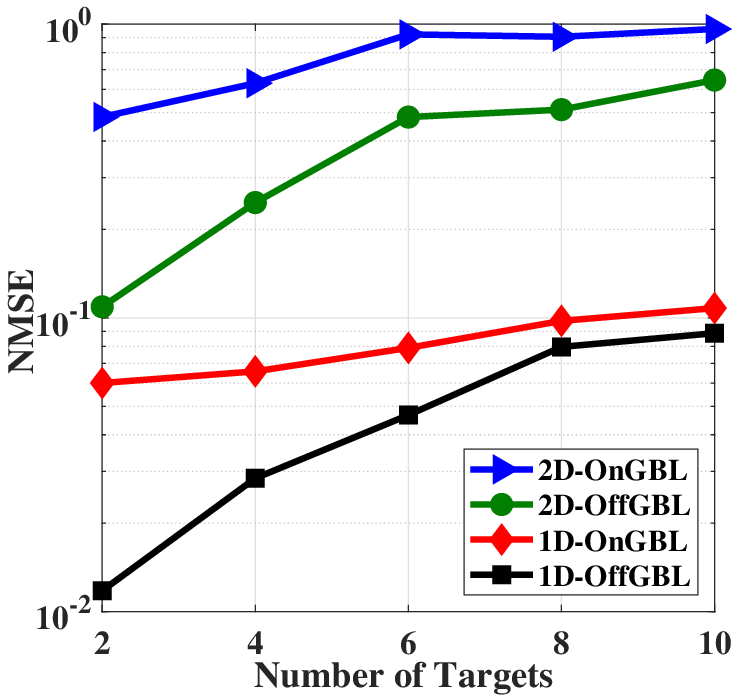}}
\caption{MIMO DCO-OFDM VL-ISAC system (a) NMSE of channel coefficient versus SNR performance; (b) NMSE of RCS coefficient versus SNR performance; (c) NMSE versus no. of targets performance at SNR $7.5$ dB.}
\label{m1}
\vspace{-2mm}
\end{figure*}
\begin{figure*}[t]
\centering
\captionsetup[subfigure]{justification=centering}
\subfloat[]{\label{nmse4}\includegraphics[width=60mm,height=50mm]{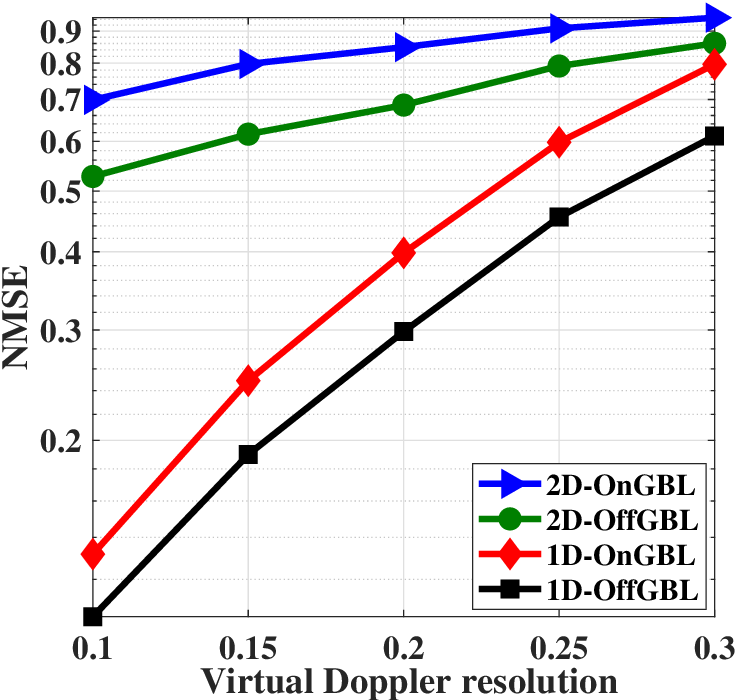}}
\subfloat[]{\label{nmse5}\includegraphics[width=60mm,height=50mm]{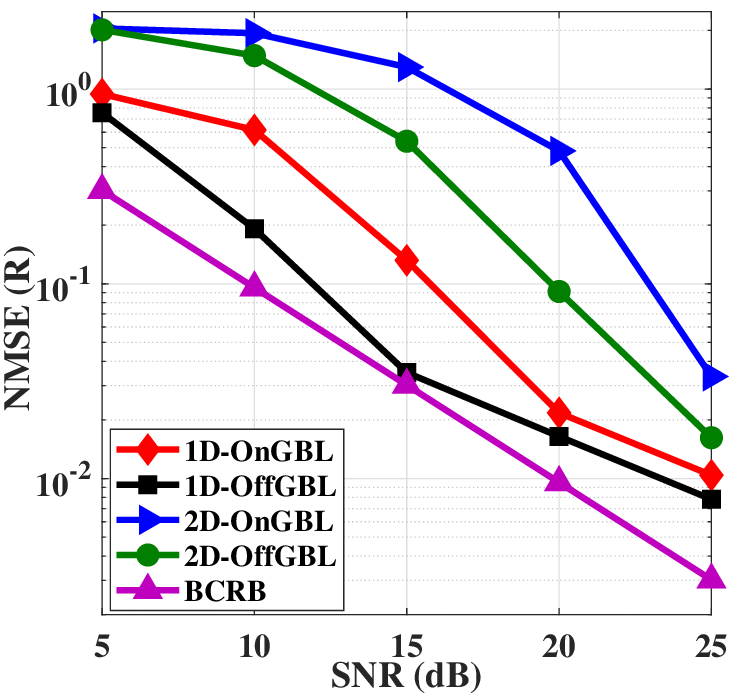}}
\subfloat[]{\label{nmse6}\includegraphics[width=60mm,height=50mm]{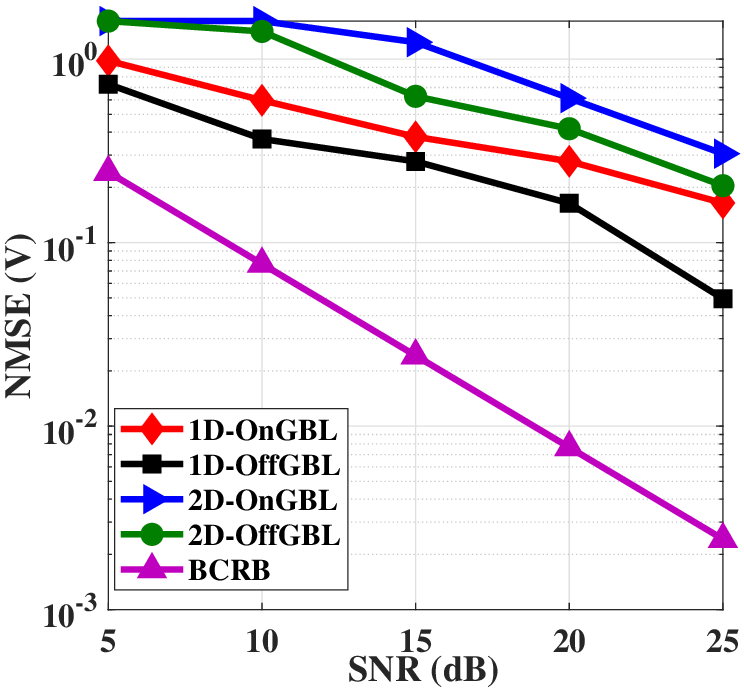}}
\caption{MIMO DCO-OFDM VL-ISAC system (a) NMSE versus Doppler resolution performance at SNR = $5$ dB; (b) NMSE of range versus SNR performance; (c) NMSE of velocity versus SNR performance.}
\label{m2}
\vspace{-2mm}
\end{figure*}
\begin{table}[t]
\centering
\caption{Simulation Parameters}
\label{tab:sim_params}
\resizebox{!}{!}{%
\begin{tabular}{|l|c|}
\hline
\textbf{Parameter} & \textbf{Value} \\ 
\hline

 System Bandwidth ($B$) & $20$ MHz
\\ Subcarrier Spacing ($\Delta{f}$) & $1.25$ MHz  \\

No.of transmitter LEDs at BS ($N_t$) & $4$ 
\\ No.of sensing PDs at BS ($N_r$) & $4$ \\

No.of receiver PDs at the UE ($R$) & $2$ 
\\ Modulation Scheme & DCO-OFDM \\

DC Bias of DCO-OFDM  & $7~\mathrm{dB}$ 
\\ Constellation Scheme & $16$-QAM \\

Number of subcarriers ($M$) & $16$ 
\\ Length of the cyclic prefix ($N_{\mathrm{CP}}$) & $4$ \\

No. of scatterers ($L$) & $5$ 
\\ No. of targets, clutters ($P,N_\mathrm{cl}$) & $4,1$ \\

Delay bins for sensing ($D_{\tau}$) & $16$ 
\\ Doppler bins for sensing ($D_{\nu}$) & $8$ \\

No. of DCO-OFDM symbols ($N$) & $8$ 
\\ Monte-Carlo simulations & $100$ \\

\hline
\end{tabular}%
}
\vspace{-3mm}
\end{table}
The performance of the proposed framework is evaluated in terms of CSI and RCS estimation accuracy. The considered schemes include $1$D on-grid BL ($1$D-OnGBL), $1$D-OffGBL, $2$D on-grid BL ($2$D-OnGBL), and $2$D-OffGBL. These methods are compared with conventional benchmarks, namely the LMMSE estimator and OMP. The estimation accuracy is evaluated using the normalized MSE (NMSE).

From the CE results in Fig. \ref{m1}\subref{nmse1}, it is observed that the $1$D-OffGBL method consistently achieves the lowest NMSE over the considered SNR range. Compared with the $1$D-OnGBL scheme, the off-grid formulation provides a noticeable gain, particularly in the high-SNR regime, due to its ability to compensate for grid mismatch caused by continuous-valued channel parameters. The $2$D BL-based methods, including both on-grid and off-grid variants, generally exhibit lower CE accuracy than their $1$D counterparts. This degradation is mainly caused by error propagation between the two estimation stages, where errors in one dimension affect the accuracy of the other dimension. OMP and LMMSE yield inferior NMSE performance because of sequential support selection and grid discretization in OMP, and the inability of LMMSE to exploit sparse VLC multipath characteristics.

A similar trend is observed for RCS estimation in Fig. \ref{m1}\subref{nmse2}. The proposed off-grid BL-based schemes achieve lower NMSE than their on-grid counterparts, which confirms their effectiveness in estimating continuous-valued RCS coefficients. In particular, the $1$D-OffGBL method provides the most accurate RCS estimates due to its refined off-grid modeling capability. The $2$D BL-based RCS estimators show slightly degraded performance compared with the $1$D methods because of inter-dimensional error propagation. However, they remain more robust to grid mismatch when coarse virtual sampling resolutions are employed. In contrast, OMP and LMMSE exhibit significantly higher NMSE values. This is because OMP is limited by greedy atom selection and finite dictionary resolution, while LMMSE does not exploit the sparsity of dominant reflective paths in the VL-based radar environment.

% Fig. \ref{m1}\subref{nmse3} shows the RCS estimation performance with respect to the number of targets. As the target density increases, a clear performance gap appears among the considered estimators. The proposed $1$D-OffGBL method maintains a low and stable NMSE, as compared to other proposed methods. This behavior is mainly due to severe error propagation across the coupled delay-Doppler dimensions, which weakens the effective sparsity in the joint $2$D domain and leads to unstable BL updates. In contrast, 
% The $1$D-OffGBL approach avoids multidimensional error amplification and provides more reliable RCS estimation as the number of targets increases. 

Fig. \ref{m1}\subref{nmse3} presents the RCS estimation performance as a function of the number of targets. As the target density increases, the performance disparity among the considered estimators becomes more pronounced. Among the proposed methods, $1$D-OffGBL maintains a consistently low NMSE by jointly exploiting the coupled DD structure. In contrast, the sequential processing employed by the $2$D methods may propagate estimation errors between the Doppler and delay stages, thereby degrading the effective sparsity and stability of the BL updates. Consequently, $1$D-OffGBL provides more reliable RCS estimation in densely populated target scenarios.

\begin{figure*}[h]
%\centering
%\captionsetup[subfigure]{justification=centering}
\hspace{-20mm}
{\includegraphics[width=200mm,height=60mm]{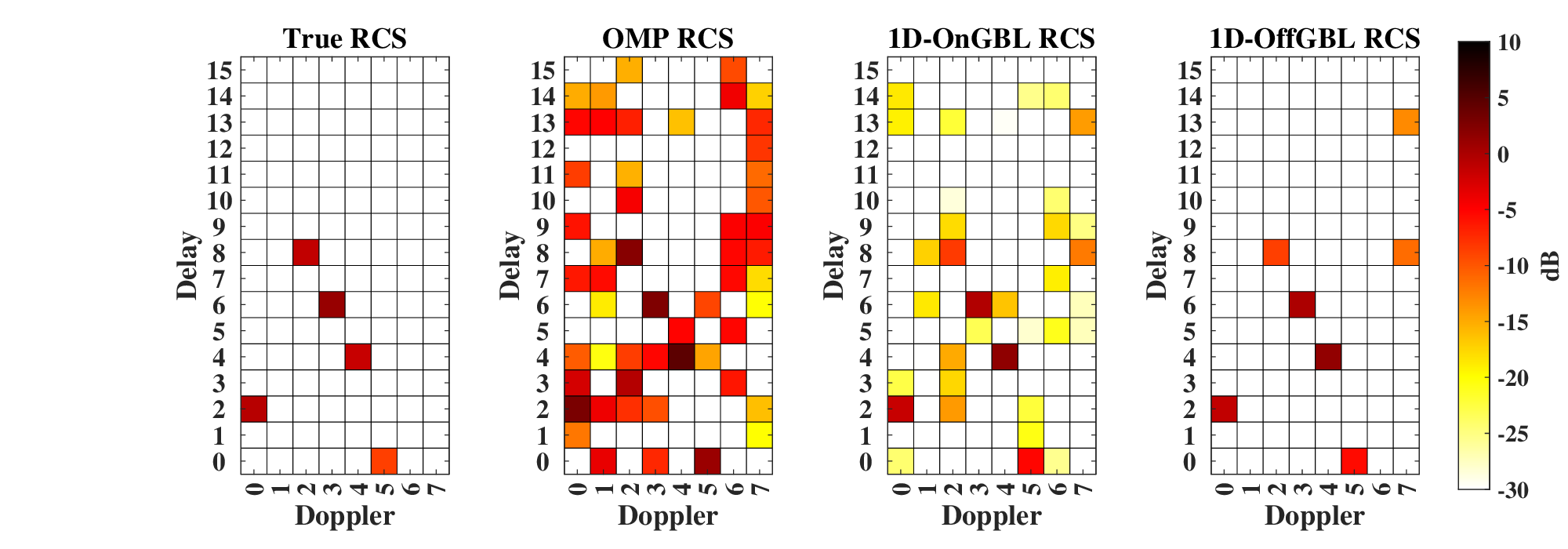}}
\caption{Heat map of RCS estimation at SNR $5$ dB for different algorithms for MIMO DCO-OFDM VL-ISAC systems.}
\label{m3}
\vspace{-2mm}
\end{figure*}

% The impact of virtual grid resolution is illustrated in Fig. \ref{m2}\subref{nmse4}. It is observed that the NMSE of $1$D-OffGBL the on-grid and off-grid BL methods intersect as the grid resolution varies.%For finer grids, the on-grid BL methods may perform better due to their simpler linear model and smaller number of unknown parameters. However, as the grid becomes coarser, the performance of the on-grid methods deteriorates because the estimated channel and RCS parameters are constrained to lie on the predefined grid. 
% In contrast, the $1$D-OffGBL methods explicitly estimate the grid deviations as additional hyperparameters, which improves both channel and RCS estimation accuracy under coarse grid settings. %The $2$D off-grid BL method provides a useful trade-off between estimation performance and computational complexity.

Fig. \ref{m2}\subref{nmse4} illustrates the NMSE performance of the proposed algorithms for different virtual Doppler resolutions. The NMSE increases monotonically with the resolution value because a coarser Doppler grid introduces greater quantization error and basis mismatch. Among the considered methods, $1$D-OffGBL consistently achieves the lowest NMSE by jointly exploiting the coupled DD structure and refining the off-grid deviations. $1$D-OnGBL exhibits moderate degradation because its estimates remain constrained to predefined grid points. The $2$D methods yield comparatively higher NMSE owing to sequential Doppler-delay processing and associated error propagation. Nevertheless, $2$D-OffGBL consistently outperforms $2$D-OnGBL, confirming the effectiveness of off-grid refinement.

% The range and velocity estimation performance is evaluated after obtaining the estimated RCS coefficient vector $\widehat{\mathbf{p}}_{\alpha}$ using the proposed framework. A target is declared to be present in a delay-velocity bin $(i,j)$ according to the detection criterion described in Section V. Fig. \ref{m2}\subref{nmse5} and \ref{m2}\subref{nmse6} show the NMSE performance of the proposed method for range and velocity estimation, with the detection threshold set to $\eta=0.01$. The estimated range-velocity map successfully identifies all dominant targets, although some false detections are observed. These false positives generally correspond to weak RCS magnitudes. As the SNR increases, the false positive rate decreases significantly, and the target range and velocity are estimated with high accuracy. Overall, the simulation results demonstrate that the proposed estimation framework can accurately recover the RCS coefficients as well as the corresponding range and velocity information of multiple targets. The results also show that the proposed approach remains effective in challenging low-SNR scenarios and provides improved detection reliability at higher SNR values. %Nevertheless, the $2$D formulations are more robust to coarse grid resolutions since the channel parameters are estimated in a decoupled manner.

Fig. \ref{m2}\subref{nmse5} and Fig. \ref{m2}\subref{nmse6} present the NMSE performance of range and velocity estimation, respectively, with the detection threshold set to $\eta=0.01$. For all the considered algorithms, the NMSE decreases consistently with increasing SNR because the enhanced signal quality enables more accurate recovery of the RCS coefficients and their corresponding DD indices. The off-grid algorithms outperform their on-grid counterparts by refining the continuous-valued delay and Doppler deviations, thereby mitigating basis mismatch. In particular, \(1\)D-OffGBL achieves the lowest NMSE among the proposed methods for both range and velocity estimation and its performance remains near the corresponding BCRB. By contrast, the \(2\)D algorithms exhibit comparatively higher NMSE because sequential Doppler and delay processing may introduce error propagation between the two estimation stages. Nevertheless, \(2\)D-OffGBL consistently improves upon \(2\)D-OnGBL. These results confirm the effectiveness of the proposed framework for accurate multi-target range and velocity estimation.

The RCS heat maps presented in Fig. \ref{m3} compare the reconstruction performance of OMP, \(1\)D-OnGBL, and \(1\)D-OffGBL with the true target distribution in the DD domain. OMP exhibits substantial energy dispersion and numerous spurious components across the DD grid. This behavior results from its greedy support selection and sensitivity to dictionary coherence and finite grid resolution. The \(1\)D-OnGBL estimator provides a more concentrated reconstruction and successfully identifies the dominant target locations. However, several low-magnitude artifacts remain because the on-grid formulation restricts the estimated delay and Doppler parameters to predefined grid points, making it susceptible to basis mismatch. In contrast, \(1\)D-OffGBL accurately reconstructs all dominant targets with sharper localization and considerably lower energy leakage. This improvement arises from the iterative refinement of the DD grid deviations, which accommodates continuous-valued target locations beyond the nominal DD grid. Although a few weak false components remain, their RCS magnitudes are substantially lower than those of the dominant targets and can be suppressed through appropriate thresholding. These observations corroborate the NMSE results. %These observations are consistent with the corresponding NMSE results.

\vspace{-2mm}
\section{Conclusions}
In this paper, BL-based PE and CE were investigated for MIMO DCO-OFDM-aided VL-ISAC systems under both on-grid and off-grid formulations. The proposed $1$D-OffGBL algorithm employs a joint DD dictionary to preserve the coupled DD structure and jointly refine the corresponding grid mismatches, thereby achieving high estimation accuracy. However, its large dictionary dimension results in increased computational complexity. To address this limitation, the proposed $2$D-OffGBL algorithm decomposes the estimation problem into sequential Doppler and delay estimation stages, avoiding the construction of a high-dimensional joint DD dictionary. Off-grid refinement was incorporated into both stages to improve the estimation of continuous-valued sensing parameters. The resulting complexity reduction, however, is accompanied by a moderate performance loss due to the decoupled processing structure. Sparse channel representations and corresponding CE procedures were also developed for both formulations, together with methods for recovering the target range and velocity. Furthermore, the derived BCRBs established theoretical benchmarks for PE, CE, range, and velocity estimation. Simulation results demonstrated that $1$D-OffGBL provides superior estimation accuracy, whereas $2$D-OffGBL offers a favorable balance between computational complexity and estimation performance. These findings confirm the flexibility of the proposed Bayesian framework for practical MIMO DCO-OFDM VL-ISAC systems.

\vspace{-2mm}
\appendix
%\section{Derivation of the Off-Grid BL Update Rules}

%This appendix presents the derivations of the update rules for the precision hyperparameter $z_g$ and the off-grid deviation vectors $\boldsymbol{\epsilon}_{\tau}$ and $\boldsymbol{\epsilon}_{\nu}$ used in the proposed $1$D off-grid BL-based radar PE algorithm.

For the update of $z_g$, the surrogate objective in \eqref{eq:block_mm_updates}, after discarding terms independent of $\mathbf{z}$, can be written as
%\vspace{-2mm}
\begin{align}
\mathcal{L}(\mathbf{z})
=
\sum_{g=0}^{G-1}
\left[
(a+N_r)\ln z_g
-
z_g
\left(
b+
\sum_{r=1}^{N_r}
\left[
\boldsymbol{\Upsilon}_r
\right]_{g,g}
\right)
\right]
+
\tilde{c}_1,
\label{eq:appendix_z_objective}
\end{align}
where $\tilde{c}_1$ is independent of $\mathbf{z}$, $G=D_\tau D_\nu N_t$, and
$\boldsymbol{\Upsilon}_r=\boldsymbol{\mu}_r\boldsymbol{\mu}_r^H+\boldsymbol{\Sigma}$. Here, $\boldsymbol{\mu}_r$ and $\boldsymbol{\Sigma}$ denote the posterior mean and covariance of $\boldsymbol{\alpha}_r$, respectively. Differentiating \eqref{eq:appendix_z_objective} with respect to $z_g$ and setting the result to zero gives
\vspace{-2mm}
\begin{align}
z_g^{(k+1)}
=
\frac{a+N_r}
{
b+
\sum_{r=1}^{N_r}
\left[
\boldsymbol{\mu}_r\boldsymbol{\mu}_r^H
+
\boldsymbol{\Sigma}
\right]_{g,g}
},
\quad
0\le g\le G-1.
\label{eq:appendix_z_update}
\end{align}

Next, the update of $\boldsymbol{\epsilon}_{\tau}$ is derived by retaining only the terms in the surrogate objective that depend on $\boldsymbol{\epsilon}_{\tau}$. For fixed $\mathbf{z}^{(k+1)}$ and $\boldsymbol{\epsilon}_{\nu}^{(k)}$, the effective dictionary can be expressed as
$\widetilde{\boldsymbol{\Phi}}_{\mathrm{MIMO}}
=
\boldsymbol{\Psi}_{\mathrm{MIMO},\tau}
+
\boldsymbol{\Phi}_{\mathrm{MIMO},\tau}
\operatorname{diag}(\boldsymbol{\epsilon}_{\tau})$,
where $\boldsymbol{\Psi}_{\mathrm{MIMO},\tau}$ contains the components independent of $\boldsymbol{\epsilon}_{\tau}$. Taking expectation with respect to the posterior distribution of $\mathbf{A}$, the relevant part of the objective becomes 
%\vspace{-3mm}
\begin{align}
\mathcal{L}(\boldsymbol{\epsilon}_{\tau})
=
-
\boldsymbol{\epsilon}_{\tau}^{H}
\mathbf{M}_{\tau}
\boldsymbol{\epsilon}_{\tau}
+
2\Re\left\{
\mathbf{p}_{\tau}^{H}
\boldsymbol{\epsilon}_{\tau}
\right\}
+
\tilde{c}_2,
\label{eq:appendix_eps_tau_objective}
\end{align}
where $\mathbf{M}_{\tau}
=
\Re\left\{
\boldsymbol{\Phi}_{\mathrm{MIMO},\tau}^{H}
\boldsymbol{\Phi}_{\mathrm{MIMO},\tau}
\odot
\left(
\mathbf{U}\mathbf{U}^{H}
+
N_r\boldsymbol{\Sigma}
\right)
\right\},$ $\mathbf{p}_{\tau}
=
\Re\left\{
\sum_{r=1}^{N_r}
\operatorname{diag}(\boldsymbol{\mu}_r)
\boldsymbol{\Phi}_{\mathrm{MIMO},\tau}^{H}
\left(
\mathbf{y}_r
-
\boldsymbol{\Psi}_{\mathrm{MIMO},\tau}
\boldsymbol{\mu}_r
\right)
\right\}
-
N_r
\Re\left\{
\operatorname{diag}
\left(
\boldsymbol{\Phi}_{\mathrm{MIMO},\tau}^{H}
\boldsymbol{\Psi}_{\mathrm{MIMO},\tau}
\boldsymbol{\Sigma}
\right)
\right\}$, and $\mathbf{U}=[\boldsymbol{\mu}_1,\boldsymbol{\mu}_2,\ldots,\boldsymbol{\mu}_{N_r}]$.
%Here, $\odot$ denotes the Hadamard product, $\Re\{\cdot\}$ extracts the real part, and $\mathbf{U}=[\boldsymbol{\mu}_1,\boldsymbol{\mu}_2,\ldots,\boldsymbol{\mu}_{N_r}]$. 
Differentiating \eqref{eq:appendix_eps_tau_objective} with respect to $\boldsymbol{\epsilon}_{\tau}$ and equating the result to zero yields
\vspace{-2mm}
\begin{align}
\boldsymbol{\epsilon}_{\tau}^{(k+1)}
=
\mathbf{M}_{\tau}^{\dagger}
\mathbf{p}_{\tau}.
\label{eq:appendix_eps_tau_update}
\end{align}
%where $(\cdot)^{\dagger}$ denotes the Moore-Penrose pseudo-inverse.

Similarly, for the update of $\boldsymbol{\epsilon}_{\nu}$, with $\mathbf{z}^{(k+1)}$ and $\boldsymbol{\epsilon}_{\tau}^{(k+1)}$ fixed, the effective dictionary is written as
$\widetilde{\boldsymbol{\Phi}}_{\mathrm{MIMO}}
=
\boldsymbol{\Psi}_{\mathrm{MIMO},\nu}
+
\boldsymbol{\Phi}_{\mathrm{MIMO},\nu}
\operatorname{diag}(\boldsymbol{\epsilon}_{\nu})$.
The corresponding surrogate objective is
%\vspace{-2mm}
\begin{align}
\mathcal{L}(\boldsymbol{\epsilon}_{\nu})
=
-
\boldsymbol{\epsilon}_{\nu}^{H}
\mathbf{M}_{\nu}
\boldsymbol{\epsilon}_{\nu}
+
2\Re\left\{
\mathbf{p}_{\nu}^{H}
\boldsymbol{\epsilon}_{\nu}
\right\}
+
\tilde{c}_3,
\label{eq:appendix_eps_nu_objective}
\end{align}
where $\mathbf{M}_{\nu}
=
\Re\left\{
\boldsymbol{\Phi}_{\mathrm{MIMO},\nu}^{H}
\boldsymbol{\Phi}_{\mathrm{MIMO},\nu}
\odot
\left(
\mathbf{U}\mathbf{U}^{H}
+
N_r\boldsymbol{\Sigma}
\right)
\right\},$ and $\mathbf{p}_{\nu}
=
\Re\left\{
\sum_{r=1}^{N_r}
\operatorname{diag}(\boldsymbol{\mu}_r)
\boldsymbol{\Phi}_{\mathrm{MIMO},\nu}^{H}
\left(
\mathbf{y}_r
-
\boldsymbol{\Psi}_{\mathrm{MIMO},\nu}
\boldsymbol{\mu}_r
\right)
\right\}
-
N_r
\Re\left\{
\operatorname{diag}
\left(
\boldsymbol{\Phi}_{\mathrm{MIMO},\nu}^{H}
\boldsymbol{\Psi}_{\mathrm{MIMO},\nu}
\boldsymbol{\Sigma}
\right)
\right\}.$
% \vspace{-2mm}
% \begin{align}
% \mathbf{M}_{\nu}
% =
% \Re\left\{
% \boldsymbol{\Phi}_{\mathrm{MIMO},\nu}^{H}
% \boldsymbol{\Phi}_{\mathrm{MIMO},\nu}
% \odot
% \left(
% \mathbf{U}\mathbf{U}^{H}
% +
% N_r\boldsymbol{\Sigma}
% \right)
% \right\},
% \label{eq:appendix_M_nu}
% \end{align}
%and \vspace{-3mm}
% \begin{align}
% \mathbf{p}_{\nu}
% &=
% \Re\left\{
% \sum_{r=1}^{N_r}
% \operatorname{diag}(\boldsymbol{\mu}_r)
% \boldsymbol{\Phi}_{\mathrm{MIMO},\nu}^{H}
% \left(
% \mathbf{y}_r
% -
% \boldsymbol{\Psi}_{\mathrm{MIMO},\nu}
% \boldsymbol{\mu}_r
% \right)
% \right\} \notag \\
% &\quad
% -
% N_r
% \Re\left\{
% \operatorname{diag}
% \left(
% \boldsymbol{\Phi}_{\mathrm{MIMO},\nu}^{H}
% \boldsymbol{\Psi}_{\mathrm{MIMO},\nu}
% \boldsymbol{\Sigma}
% \right)
% \right\}.
% \label{eq:appendix_p_nu}
% \end{align}
Therefore, the Doppler off-grid deviation update is given as
\vspace{-2mm}
\begin{align}
\boldsymbol{\epsilon}_{\nu}^{(k+1)}
=
\mathbf{M}_{\nu}^{\dagger}
\mathbf{p}_{\nu}.
\label{eq:appendix_eps_nu_update}
\end{align}
Equations \eqref{eq:appendix_z_update}, \eqref{eq:appendix_eps_tau_update}, and \eqref{eq:appendix_eps_nu_update} correspond to the update rules for $z_g$, $\boldsymbol{\epsilon}_{\tau}$, and $\boldsymbol{\epsilon}_{\nu}$ in \eqref{eq:z_update}, \eqref{eq:offgrid_update_eps_tau}, and \eqref{eq:offgrid_updates} respectively.
\vspace{-2mm}
\bibliographystyle{IEEEtran}
\bibliography{citation}

\end{document}